# High Efficiency Inference Accelerating Algorithm for NOMA-based Mobile Edge Computing

Xin Yuan, Ning Li, *Member*, *IEEE*, Tuo Zhang, Muqing Li, Yuwen Chen, Jose Fernan Martinez Ortega, Song Guo, *Fellow, IEEE*

*Abstract*—Splitting the inference model between device, edge server, and cloud can improve the performance of EI greatly. Additionally, the non-orthogonal multiple access (NOMA), which is the key supporting technologies of B5G/6G, can achieve massive connections and high spectrum efficiency. Motivated by the benefits of NOMA, integrating NOMA with model split in MEC to reduce the inference latency further becomes attractive. However, the NOMA based communication during split inference has not been properly considered in previous works. Therefore, in this paper, we integrate the NOMA into split inference in MEC, and propose the effective communication and computing resource allocation algorithm to accelerate the model inference at edge. Specifically, when the mobile user has a large model inference task needed to be calculated in the NOMA-based MEC, it will take the energy consumption of both device and edge server and the inference latency into account to find the optimal model split strategy, subchannel allocation strategy (uplink and downlink), and transmission power allocation strategy (uplink and downlink). Since the minimum inference delay and energy consumption cannot be satisfied simultaneously, and the variables of subchannel allocation and model split are discrete, the gradient descent (GD) algorithm is adopted to find the optimal tradeoff between them. Moreover, the loop iteration GD approach (Li-GD) is proposed to reduce the complexity of GD algorithm that caused by the parameter discrete. Additionally, the properties of the proposed algorithm are also investigated, which demonstrate the effectiveness of the proposed algorithms.

## I. INTRODUCTION

The artificial intelligence has been widely used and changed our life greatly, such as metaverse [1-2], automatic driving [2-4], image generation [5], etc. However, since the AI model is always large for achieving high accuracy, the computing resource that needed for these models are huge. Therefore, it is inappropriate to deploy these AI models on the mobile devices, such as mobile phones and vehicles, in which the computing resource is quite limited. For addressing this issue, one possible solution is to divide the large AI model into different parts and offload the resource-intensive model to edge server for reducing resource requirement and latency [6-8, 51-53]. The model split between device, edge server and cloud has been investigated deeply by previous works, such as [9-16]. These works find the optimal model segmentation point and early-exist point to minimize inference delay and the allocated resource while maintain high inference accuracy by reinforcement learning [9], convex optimization [10-13], heuristic algorithm [14-16], etc. Based on these approaches, many large AI models have been deployed on resource limited devices successfully to provide high quality intelligent services.

In addition, to facilitate the massive connectivity over 5G/6G networks, non-orthogonal multiple access (NOMA) technique [17-22] has been proposed as an alternative for traditional OMA, which can achieve massive connections and high spectrum efficiency. The NOMA has been selected as one of the key supporting technologies of B5G/6G. The key idea of NOMA is to allow multiple users to transmit on the same resource block simultaneously, and successive interference cancelation (SIC) is used as an efficient multi-user detection technique at the receiver to reduce the co-channel interference and decode the target signal. Integrating NOMA with multi-access MEC has been investigated deeply, such as [23-32]. In NOMA-based MEC system, the data transmission rate of devices can be ensured through appropriate transmission power allocation based on their channel conditions [23]. In these works, the NOMA is used in uplink and downlink for task offloading and final results transmission. Then the task offloading strategy, the channel allocation strategy, and the transmission power allocation strategy are optimized jointly to achieve high spectrum efficiency, high transmission rate, and low transmission delay. Motivated by the benefits of NOMA, integrating NOMA with model split in MEC to reduce the inference latency further becomes attractive to the researchers.

However, the NOMA-based communication aspect has not been properly considered during model split inference in MEC. Even the bandwidth has been considered during model split in previous works, such as [21-25], this issue is further complicated by the sophistication in communication resource allocation caused by the NOMA scheme. Under the NOMA scheme, multiple mobile users can be served by one subchannel with ensured data rates through proper transmit power allocation based on their channel conditions. For the NOMA-based model split inference in MEC, it must ensure that the Tx Power and Rx power of mobile users are adequate to guarantee their data transmission rate. However, none of the existing model split inference has considered these unique characteristics of NOMA. Moreover, the data transmission rate changes with different channel allocation strategies and it has great effect on the inference performance in split learning. This is because for the DNN, as shown in [9-15], the intermediate data size varies with different layers, and the output of the shallow layers is always larger than the deep layers. Thus, when the channel transmission rate is low, it tends to split the model at deep layer and offload small part model to edge server, which causes inference delay and energy consumption of end devices; when the channel transmission rate is high, it tends to split at shallow layer, then the inference delay is reduced while the energy consumption on edge server is increased. Additionally, due to the different capabilities of device and edge server, for the same model, the model quantization accuracy on device and edge server may be different. Therefore, different channel allocation strategy and model split strategy also can affect the inference accuracy. Additionally, for the mobile device, one of the most important factors to consider is the energy consumption, because it is always powered by battery which is limited on energy store and decides the lifetime of devices. In most scenarios of MEC, the lifetime of mobile devices is more important than inference delay and accuracy to some extent, because long lifetime is the foundation of high QoS [33-35]. Once the mobile devices are out of energy, the QoS is meaningless. Therefore, the energy consumption of mobile devices is critical to edge intelligence. Except for the end devices, the edge servers running 24/7 consume a large amount of energy, contribute a significant proportion of global carbon emissions, and thus the energy consumption on edge server should be reduced to achieve green data center. Moreover, considering the transmission power of device and edge server can be adjusted in NOMA to achieve different transmission rate, they should be paid attention.

Solving the above issues is challenging. Firstly, finding the optimal model split and resource allocation strategy when considering both energy consumption and inference delay is not easy. Because the optimal objectives of these two parameters are opposite. For instance, when minimizing the energy consumption of edge server, the size of model that offloaded to edge server should be as small as possible. However, this means most part of the model should be calculated on device, which will cause high inference delay and energy consumption of end device. Therefore, how to find the optimal tradeoff between these two parameters and optimize the performance of the whole systems is difficult. Secondly, in the NOMA scheme, the device and edge server can adjust their data transmission rate by adjusting the transmission power and selecting different subchannel. And the data transmission rate has great effect on the performance of split inference and energy consumption. Therefore, the tight coupling of these parameters makes find the optimal solution is challenging. Moreover, considering that the channel allocation strategy and model split strategy are all discrete, addressing this issue becomes even more difficult.

Based on the above issues, in this paper, the Effective Communication (subchannel and transmission power) and Computing resource allocation algorithm is proposed for accelerating the split inference in NOMA-based MEC, shorted as ECC. Specifically, when the mobile user has a large model inference task needed to be calculated in the NOMA-based MEC, it will take the energy consumption of both device and edge server and the inference latency into account to find the optimal model split strategy, subchannel allocation strategy (uplink and downlink), and transmission power allocation strategy (uplink and downlink). Since the minimum inference delay and energy consumption cannot be satisfied simultaneously, and the variables of subchannel allocation and model split are discrete, the gradient descent (GD) algorithm is adopted to find the optimal tradeoff between them. Moreover, the loop iteration GD approach (Li-GD) is proposed to reduce the complexity of GD algorithm that caused by the parameter discrete. Additionally, the properties of the proposed algorithms are investigated, including convergence, complexity, and approximate error.

The contributions of this paper can be summarized as follows.

1) First, we integrate the NOMA into model split inference between device and edge server. In this approach, both the energy consumption and inference delay are considered to find the optimal model segmentation and resource allocation strategy in NOMA-based split inference. Moreover, considering the complexity of this problem, we proposed the Li-GD algorithm to achieve optimal tradeoff between inference delay and energy consumption effectively.
2) Then, the properties of the proposed Li-GD algorithm are investigated. First, it can be proved that the Li-GD algorithm is convergent, and the convergence time is $K = \frac{\|x^0 - x^*\|_2^2}{2\eta\epsilon}$, the complexity of the Li-GD is $O(X\overline{K}\mathcal{F}Mx^3 \ln^2(x))$, the approximate error is smaller than $\frac{\varepsilon}{\rho_{min}(1-B_{max})\log_2\left(1+\frac{P_{min}}{\Delta^* + \frac{\alpha P_{max}}{2}}\right)}$. Additionally, it can be demonstrated that the Li-GD algorithm can reduce complexity and convergence time compared to traditional GD approach.

The rest of this paper is organized as follows. Section II introduces the related works. The network models and the problems that will be solved in this paper is presented in Section III. The Li-GD algorithm is proposed in Section IV, and the properties, e.g., the convergence, the complexity, etc., are also investigated in this section. In Section V, the effectiveness of the proposed ECC algorithm is demonstrated by simulation. Section VI summarizes the conclusions of this work.

## II. RELATED WORKS

### A. NOMA-based mobile edge computing

During task offloading and results transmission in MEC, the NOMA has been used in uplink or downlink channel. Then multi-objectives, such as the task offloading strategy, and channel allocation strategy, etc., are optimized jointly to achieve high spectrum efficiency, high transmission rate, and low transmission delay of MEC.

Specifically, the authors in [23] proposed heuristic algorithms for MEC-aware NOMA single cell, in which the user clustering, computation resource and transmission powers allocation were jointly optimized to minimize the energy consumption. In [24], the authors proposed a joint optimization scheme of SIC ordering and computation resource in a MEC-aware NOMA NB-IoT network, in order to minimize the total execution delay for each bit. The authors in [25] considered two kinds of users' scenarios, including the single user and the multiple users, in which the offloading workloads and transmission time were both optimized to minimize the delay, while satisfying the computation requirement of each user. In [26], the authors proposed a scheduling-based algorithm for the device-to-device (D2D)-assisted and NOMA-based MEC system, in which D2D was applied to enhance the user collaboration and relieve the computing burden on edge server, and the computation resource, power, and channel allocations were jointly optimized to minimize the system cost. The aforementioned works focus on the single-cell networks. The NOMA-assisted MEC in multi-cell networks has been learned in [27-32]. Specifically, the authors in [27] investigated the joint power and computation resource allocation scheme for the NOMA-assisted MEC heterogenous network (HetNet), to minimize the energy consumption of users. In [28], the authors proposed an energy-efficient joint offloading decision, user scheduling and resource allocation scheme for an integrated fog-cloud network. Moreover, [29] investigate the online user allocation problem in MEC with NOMA. The authors minimize the allocation delay and transmit power costs by Lyapunov to convert the long-term optimization problem into a series of subproblems in every time slot. The authors in [30] propose an online algorithm to solve the dynamic edge user allocation problem in NOMA-based MEC system. The authors in [31] tackle the edge demand response problem in NOMA-based MEC system. A two-phase game-theoretical approach is proposed in [32].

*B. Model split and resource allocation*

For the model splitting and resource allocation, the re-search efforts focus on offloading computation from the resource-constrained mobile to the powerful cloud to reduce inference time [36][37]. Neurosurgeon [38] explores a computation offloading method for DNNs be-tween the mobile device and the cloud server at layer granularity. In [39], the authors propose MAUI, which is an offloading framework that can determine to execute the functions of a program on edge or cloud. But the MAUI is not explicitly designed for DNN segmentation since the size of communication data between devices and edge/cloud is small. The DDNN is proposed in [40], which is a distributed deep neural network architecture that is distributed across different computing devices. The purpose of DDN is to reduce the communication data size among devices for the given DNN. The authors in [9] pro-pose multi-exit DNN inference acceleration framework based on multi-dimensional optimization. This is the first work that investigates the bottlenecks of executing multi-exit DNNs in edge computing and builds a novel model for inference acceleration with exit selection, model partition, and resource allocation. The authors of [41] and [42] propose feature-based slicing method, which divided a layer into multiple slices for parallel inference. The [41] established a mobile devices' cluster and published the slices to appropriate nodes. While the [42] developed adaptive partitioning and offloading of multi-end devices to multi-edge nodes based on a matching game approach. Unlike slicing feature map, the authors of [43] sliced the input data into multiple tiles for parallelism. However, both feature-based and input-based partition require frequent synchronous data transmission. Other studies adopt layer wise partition. The authors of [44] extended to the multi-device environment and proposed an iterative algorithm for resource allocation. However, in most cases, model partition degenerates into binary offloading. Thus, the authors of [45] and [46] compressed the intermediate data to reduce the transmission latency with a certain loss of accuracy. In [47], by modeling and solving a batch task scheduling problem, the feasibility of partitioning on a three-exit DNN is also verified under a three-tier framework, i.e., devices, edge, and cloud. In [10], the authors devise a collaborative edge computing system CoopAI to distribute DNN inference over several edge devices with a novel model partition technique to allow the edge devices to prefetch the required data in advance to compute the inference cooperatively in parallel without exchanging data. In [11], the authors propose a technique to divide a DNN in multiple partitions that can be processed locally by end devices or offloaded to one or multiple powerful nodes, such as in fog networks. In [12], the authors propose JointDNN, for collaborative computation between a mobile device and cloud for DNNs in both inference and training phase. In [13], considering online exit prediction and model execution optimization for multi-exit DNN, the authors propose a dynamic path based DNN synergistic inference acceleration framework (DPDS), in which the exit designators are de-signed to avoid iterative entry for exits. Moreover, the multi-exit DNN is dynamically partitioned according to network environment to achieve fine-grained computing offloading. In [14], the authors design the DNN surgery. The DNN surgery allows partitioned DNN to be processed at both the edge and cloud while limiting the data transmission. Moreover, they design a dynamic adaptive DNN surgery (DADS) scheme to optimally partitions the DNN under different network conditions. In [15], the authors investigate the optimization problem of DNN partitioning in a realistic multiuser resource-constrained condition that rarely considered in previous works. And they propose iterative alternating optimization (IAO) algorithm to achieve the optimal solution in polynomial time.

## III. NETWORK MODEL AND PROBLEM STATEMENT

In this section, the inference delay model and the energy consumption model that used in this paper are introduced. The network model is presented in Fig.1. Then, based on the proposed models, the problems that will be solved in this paper are described in detail.

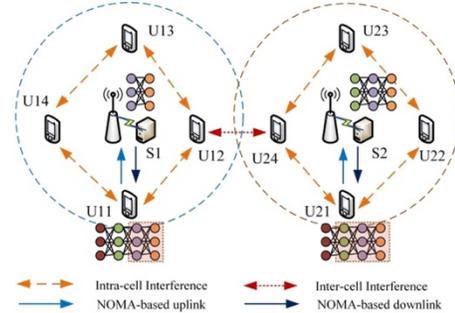

Fig. 1. Network model.

As shown in Fig.1, we consider mobile edge computing in a multi-cell network with $N$ single-antenna APs and $U$ single-antenna end devices, indexed by $N = \{1,2,\ldots,N\}$ and $U = \{1,2,\ldots,U\}$, respectively. The total system bandwidth $B$ is equally divided into $M$ orthogonal subchannels, indexed by $M = \{1,2,\ldots,M\}$. We consider the use of the nearest AP association policy [48]. As such, each end device associates with the nearest AP that can provide the maximum average channel gain. We denote the set of devices served by AP $n$ as $U_n$. Therefore, each device $u \in U$ offloads its split inference model to its associated AP $n \in N$ via subchannel $m \in M$. We assume that $M < N$ and denote the set of devices served by AP $n$ on subchannel $m$ as $U_n^m$. Therefore, the devices associated with different APs may access to the same subchannel and interfere with each other. The uplink and downlink between mobile user and edge server are all NOMA-based channels. Thus, not only the users in the coverage of the same AP can be interfered with each other, but also interfered by the users which selecting the same subchannel in adjacent APs. For each mobile user, there is a large model with $\mathcal{F}$ layers needs to be split and offloaded from device to edge server.

*A. Inference delay model*

The inference delay includes three parts: 1) the delay that caused by model inference on mobile device; 2) the delay that caused by model inference on edge server; 3) the delay that caused by intermediate data transmission between device and edge server.

**(1) Inference delay on device**

Let the number of model's layers is $\mathcal{F}$ and the model split decision is $s_i$, which means that the first to $s_i$-th layers are calculated on mobile device $i$, and the $(s_i + 1)$-th to $\mathcal{F}$-th layers are offloaded to the edge server for deep inference. $c_i$ is defined as the floating-point operation capability of device $i$. Then, the inference latency on mobile devices after model segmentation can be calculated as:

$$T_i^{device} = \sum_{\delta=1}^{s_i} \frac{f_{l_\delta}}{c_i} \quad (1)$$

where $f_{l_\delta}$ is the computation task of each layer in the main branch, containing convolutional layer $f_{conv}$, pooling layer $f_{pool}$, and ReLU layer $f_{relu}$ [9]. Thus, $f_{l_\delta}$ can be computed as:

$$f_{l_\delta} = m_{\delta 1} f_{conv} + m_{\delta 2} f_{pool} + m_{\delta 3} f_{relu} \quad (2)$$

where $m_{\delta 1}$, $m_{\delta 2}$, and $m_{\delta 3}$ denote the number of convolutional layers, pooling layers, and ReLU layers, respectively, and $m_{\delta 1} + m_{\delta 2} + m_{\delta 3} = s_i$.

**(2) Inference delay on edge server**

The execution time is not proportional to the amount of allocated computational resources for the inference tasks, such as DNN, under the scenario that the edge server is multicore CPU. As demonstrated in [15], up to 44% error in execution time between theory and experiment. Thus, in this paper, let $r_i$ represent the number of minimum computational resource unit that allocated to user $i$; $c_{min}$ implies the capability of minimum computational resource unit. Since in the multicore CPU scenario, the execution time is not linear with respect to the amount of allocated computational resource, a compensation function $\lambda(r_i)$ is introduced to fit the execution time in the multicore CPU scenario. For the single core scenario, the $\lambda(r_i)$ is degenerated to $r_i$, and for the multicore scenario, $\lambda(r_i) > r_i$. The $\lambda(r_i)$ can be estimated based on the approach that proposed in [15]. Therefore, in this paper, we only assume that $\lambda(r_i)$ increases with $r_i$, but not linear. To model the nonlinearity in the execution time, the execution time on edge can be expressed as:

$$T_i^{server} = \sum_{\delta=s_i+1}^{\mathcal{F}} \frac{f_{e_\delta}}{\lambda(r_i) c_{min}} \quad (3)$$

where $f_{e_\delta}$ is the computation task of each layer in the main branch, containing convolutional layer $f_{conv}$, pooling layer $f_{pool}$, and ReLU layer $f_{relu}$. Thus, $f_{e_\delta}$ can be calculated as:

$$f_{e_\delta} = m_{\delta 4} f_{conv} + m_{\delta 5} f_{pool} + m_{\delta 6} f_{relu} \quad (4)$$

where $m_{\delta 4}$, $m_{\delta 5}$, and $m_{\delta 6}$ denote the number of convolutional layers, pooling layers, and ReLU layers, respectively, and $m_{\delta 4} + m_{\delta 5} + m_{\delta 6} = \mathcal{F} - s_i$.

**(3) Network transmission delay**

There are two different kinds of network transmission delay: 1) the intermediate output transmission delay from device to edge server in uplink and 2) the final result transmission delay from edge server to device in downlink.

Firstly, for the intermediate data transmission delay, when the model is split at $s_i$-th layer, the intermediate data generated by the $s_i$-th layer will be transmitted to edge server to complete the inference. As shown in Fig.1, the end devices served by the same AP $n$ on the same subchannel $m$ form a NOMA cluster $U_n^m$, where NOMA protocol is adopted to perform the model offloading. Let $N_n^m$ and $h_{n,i}^m$ denote the number of devices in $U_n^m$ and the channel coefficient of the uplink from device $i$ to AP $n$ on subchannel $m$, respectively. Without loss of generality, we assume that $|h_{n,1}^m|^2 > |h_{n,2}^m|^2 > \cdots > |h_{n,i}^m|^2 > \cdots > |h_{n,N}^m|^2$. According to the rules of NOMA protocol, the APs apply SIC for multi-user detection. Specifically, each AP $n$ sequentially decodes the signal from devices with higher channel gains and regards all the other signals as the interference. As such, the received signal-to-interference-plus-noise ratio (SINR) of AP $n$ for device $i \in U_n^m$ is given by:

$$\Upsilon_{n,i}^m = \frac{p_{n,i}^m |h_{n,i}^m|^2}{\sum_{v=i+1}^{U} \beta_{n,v}^m p_{n,v}^m |h_{n,v}^m|^2 + \sum_{l=1,l\neq n}^{N} \sum_{t=1}^{U} \beta_{l,t}^m p_{l,t}^m |g_{l,t}^m|^2 + \sigma^2} \quad (5)$$

where $p_{n,i}^m$ is the transmission power of device $i \in U_n^m$, $g_{l,t}^m$ is the channel coefficient of the interference link from device $t$ served by AP $l$ except AP $n$ on the same subchannel $m$, $\sigma^2$ is the additive white Gaussian noise. Moreover, $\sum_{v=i+1}^{U} \beta_{n,v}^m p_{n,v}^m |h_{n,v}^m|^2$ is the intra-cell interference and $\sum_{l=1,l\neq n}^{N} \sum_{t=1}^{U} \beta_{l,t}^m p_{l,t}^m |g_{l,t}^m|^2$ is the inter-cell interference. Then, based on (5), the transmission rate of device $i \in U_n^m$ can be expressed as:

$$R_{n,i}^m = \beta_{n,i}^m \cdot \frac{B_{up}}{M} \log_2(1 + \Upsilon_{n,i}^m) \quad (6)$$

where $\beta_{n,i}^m$ represents the subchannel allocation variable. Specifically, $\beta_{n,i}^m = 1$ means that subchannel $m$ is allocated to device $i$; otherwise, $\beta_{n,i}^m = 0$. Let $w_{s_i}$ represent the data size at the $s_i$-th layer, then the intermediate output transmission delay can be calculated as:

$$T_i^{tran-i} = \frac{w_{s_i}}{R_{n,i}^m} = \frac{w_{s_i}}{\beta_{n,i}^m \frac{B_{up}}{M} \log_2(1+\Upsilon_{n,i}^m)} \quad (7)$$

For the final data transmission from edge server to user with downlink NOMA scheme, the AP transmits a superposition-coded signal to each user on a subchannel [49]. In downlink transmissions, users employ SIC to decode the received superposed signal. Without loss of generality, suppose that the devices served by the same AP $j$ on the same subchannel $k$ are represented by $U_j^k$. Let $N_j^k$ and $H_{j,i}^k$ denote the number of devices in $U_j^k$ and the channel coefficient of the downlink from edge server $j$ to device $i$. Similar to the uplink, we assume that $|H_{j,1}^k|^2 < |H_{j,2}^k|^2 < \cdots < |H_{j,i}^k|^2 < \cdots < |H_{j,N}^k|^2$. For user 1, since it has the weakest channel coefficient, it decodes the superposed signal from edge server $j$ without performing SIC. Then the user $1's$ decoded component is subtracted from the superposed signal. The subsequent user in $U_j^k$, i.e., user 2, can decode the received signal without interference from user 1. Following this principle, the signal received by user $i \in U_j^k$ on subchannel $k$ in BS $j$ has a SINR $\Psi_{j,i}^k$ of:

$$\Psi_{j,i}^k = \frac{|H_{j,i}^k|^2 P_{j,i}^k}{\sum_{q=i+1}^{U} \beta_{j,q}^k P_{j,q}^k |H_{j,q}^k|^2 + \sum_{x=1,x\neq j}^{N} \sum_{y=1}^{U} \beta_{x,y}^k P_{x,y}^k |G_{x,y}^k|^2 + \sigma^2} \quad (8)$$

where $|H_{j,i}^k|^2$ is channel gain of user $i$ on subchannel $k$,

$\sum_{q=i+1}^{U} \beta_{j,q}^{k} P_{j,q}^{k} |H_{j,q}^{k}|^2$ is the intra-cell interference experienced by user $i$, $\sum_{x=1,x\neq j}^{N} \sum_{y=1}^{U} \beta_{x,y}^{k} P_{x,y}^{k} |G_{x,y}^{k}|^2$ is the inter-cell interference experienced by user $i$ (caused by the neighbor AP of users $i$), and $\sigma^2$ is the addictive white Gaussian noise. In the downlink NOMA, considering the factors that affect the condition of subchannel, the SIC decoding order of users on subchannel $j$ must be the weaker users (high inter-cell interference and low channel gain) decode before stronger users. According to [50], the achievable data rate of user $i$ by SIC can be expressed as:

$$\Phi_{j,i}^{k} = \beta_{j,i}^{k} \frac{B_{down}}{M} \log_2(1 + \Psi_{j,i}^{k}) \quad (9)$$

Let $m_i$ denotes the data size of the final inference result at edge server, then the final result transmission delay can be expressed by:

$$T_i^{tran-f} = \frac{m_i}{\Phi_i^k} = \frac{m_i}{\beta_{j,i}^{k} \frac{B_{down}}{M} \log_2(1+\Psi_{j,i}^{k})} \quad (10)$$

Thus, the network transmission delay can be expressed as:

$$T_i^{trans} = T_i^{tran-i} + T_i^{tran-f} = \frac{w_{s_i}}{R_{n,i}^m} + \frac{m_i}{\Phi_{j,i}^k} \quad (11)$$

Intuitively, the overall execution latency of the task in mobile user $i$ can be expressed as:

$$T_i = T_i^{device} + T_i^{server} + T_i^{trans}$$
$$= \sum_{\delta=1}^{s_i} \frac{f l_\delta}{c_i} + \sum_{\delta=s_i+1}^{\mathcal{F}} \frac{f e_\delta}{\lambda(r_i)c_{min}} + \frac{w_{s_i}}{R_{n,i}^m} + \frac{m_i}{\Phi_{j,i}^k} \quad (12)$$

*B. Energy consumption model*

The energy consumption that considered in this paper includes: 1) the energy that consumed for model inference at end device and edge server, and 2) the energy that consumed for intermediate data transmission between end device and edge server.

**(1) Energy consumption on device**

Let $\xi_i$ represent the effective switched capacitance of CPU, which is determined by the chip structure of mobile device $i$, then the energy consumption of mobile devices that caused for model inference can be calculated as:

$$E_i^l = \sum_{\delta=1}^{s_i} \xi_i c_i^2 \varphi_i f l_\delta \quad (13)$$

where $\varphi_i$ is the required CPU cycles to compute 1-bit data on end device.

Let $p_{n,i}^m$ denotes the transmission power of mobile device $i$ on subchannel $m$, the energy consumption of intermediate data transmission can be computed as:

$$E_i^t = p_{n,i}^m \cdot \frac{w_{s_i}}{R_{n,i}^m} \quad (14)$$

**(2) Energy consumption on edge server**

For edge server, its energy consumption also comes from two aspects: the model inference and final result transmission. Based on (10), the energy consumption of intermediate data transmission can be computed as:

$$E_e^t = P_{j,i}^k \cdot \frac{m_i}{\Phi_{j,i}^k} \quad (15)$$

Let $\xi_e$ represent the effective switched capacitance of CPU, which is determined by the chip structure of edge server $e \in N$, then the energy consumption of edge server that used for model inference can be calculated as:

$$E_e^l = \sum_{\delta=s_i+1}^{\mathcal{F}} \xi_e (\lambda(r_i)c_{min})^2 \varphi_e f l_\delta \quad (16)$$

where $\varphi_e$ is the required CPU cycles to compute 1-bit data on edge server.

Thus, the energy consumption for task execution and intermediate data transmission between mobile device and edge server can be derived by:

$$E_i = E_i^l + E_i^t + E_e^l + E_e^t$$
$$= \sum_{\delta=1}^{s_i} \xi_i c_i^2 \varphi_i f l_\delta + \sum_{\delta=s_i+1}^{\mathcal{F}} \xi_e (\lambda(r_i)c_{min})^2 \varphi_e f l_\delta$$
$$+ p_{n,i}^m \cdot \frac{w_{s_i}}{R_{n,i}^m} + P_{j,i}^k \cdot \frac{m_i}{\Phi_{j,i}^k} \quad (17)$$

*C. Problem statement*

From the above subsections, we get the inference delay $T_i(s_i, \beta_{n,i}^m, \beta_{j,i}^k, p_{n,i}^m, P_{j,i}^k, r_i)$ and the energy consumption $E_i(s_i, \beta_{n,i}^m, \beta_{j,i}^k, p_{n,i}^m, P_{j,i}^k, r_i)$. Our purpose is to achieve minimum inference delay and energy consumption at the same time, with the variables are the model split strategy $\mathcal{F} = \{0,1,2,...,s_i,...,\mathcal{F}\}$, the subchannel allocation strategy of user $i$ $B_i = \{\beta_{n,i}^m, \beta_{j,i}^k\}$, the computing resource allocation strategy $r_i \in [r_{min}\ r_{max}]$, and the transmission power allocation strategy of device $i$ $P_i = \{p_{n,i}^m, P_{j,i}^k\}$. During these variables, $\mathcal{F}$ and $B_i$ are discrete, while $P_i$ and $r_i$ are continuous. Therefore, the problem (P0) that to be solved in this paper can be expressed as:

$$\min\{\sum_{i=1}^{U} T_i(s_i, B_i, P_i, r_i), \sum_{i=1}^{U} E_i(s_i, B_i, P_i, r_i)\} \quad (18)$$

$$s.t. \quad 0 \leq s_i \leq \mathcal{F}, \forall i \in [1\ U] \quad (18.a)$$
$$B_i \in \{0,1\} \quad (18.b)$$
$$p_{min} \leq P_i \leq p_{max}, \forall i \in [1\ U] \quad (18.c)$$
$$r_{min} \leq r_i \leq r_{max}, \forall i \in [1\ U] \quad (18.d)$$
$$\sum_{n=1}^{N} \sum_{m=1}^{M} \beta_{n,i}^m = 1, \forall i \in [1\ U] \quad (18.e)$$
$$\sum_{j=1}^{N} \sum_{k=1}^{M} \beta_{j,i}^k = 1, \forall i \in [1\ U] \quad (18.f)$$

In P0, the constraints of (18.a), (18.b), (18.c), and (18.d) are easy to be understood; the constraints (18.e) and (18.f) mean that each user can only select one edge server and one subchannel in one time slot. The P0 is difficult to be dealt with, since these two optimal objectives are opposite, which has been demonstrated in Section I. This indicates that finding the minimum energy consumption and minimum inference delay simultaneously is impossible. Thus, in this paper, we need to find an approach to achieve optimal tradeoff between these two objectives.

## IV. OPTIMAL MODEL SPLIT AND RESOURCE ALLOCATION ALGORITHM

In this section, we will investigate the optimal solution for P0. Since P0 is difficult to be solved, we propose an approximate algorithm for P0 in this section. Moreover, the properties of the proposed algorithm are also investigated in this section.

*A. Loop iteration GD algorithm*

Since the optimization objectives shown in P0 are opposite, we introduce the weight-based approach to construct the utility function for each mobile user that contains both these two objectives, which can be expressed as:

$$U_i = \omega_T T_i + \omega_E E_i \quad (19)$$

where $\omega_T$ and $\omega_E$ are the weights of inference delay and energy consumption, respectively, and $\omega_T + \omega_E = 1$. The weight represents the importance of each optimal objective to users.

Additionally, the $\omega_T$ and $\omega_E$ are hyper-parameters, which can be decided by mobile users according to their dynamic QoS requirements. For instance, if the inference delay is more important to mobile user than energy consumption, then the mobile user can set $\omega_T > \omega_E$. This approach is flexible and practicable due to the following reasons. First, in practice, the QoS requirements of the same mobile user may change according to the dynamic environment. For instance, when the mobile devices have enough energy, the inference delay may be the primary factor to be considered; otherwise, when the application requires low inference latency, no matter whether the energy consumption is high or not, the weight of inference delay should be large. Second, for various mobile users and applications, their requirements on QoS may be different. For instance, the energy supply in user $A$ is more than that in user $B$, then the weigh of $\omega_E$ in user $A$ may be smaller than that in user $B$; if one application has strict restriction on inference delay, then the weight of $\omega_T$ in this user could be larger than the other users. Thus, this approach can be adjusted according to the dynamic QoS requirements flexibly.

According to (19), the P0 can be expressed as:
$$min \sum_{i=1}^{U} U_i \quad (20)$$

Let $\Gamma = \sum_{i=1}^{U} U_i$, based on (8), (12) and (16), $\Gamma$ can be described as:

$$\Gamma = \sum_{i=1}^{U} \omega_T T_i(s_i, B_i, P_i, r_i) + \sum_{i=1}^{U} \omega_E^i E_i(s_i, B_i, P_i, r_i)$$
$$= \sum_{i=1}^{U} \omega_T^i \left( \sum_{\delta=1}^{s_i} \frac{f_{l_\delta}}{c_i} + \sum_{\delta=s_i+1}^{\mathcal{F}} \frac{f_{e_\delta}}{\lambda(r_i)c_{min}} + \frac{w_{s_i}}{\beta_{n,i}^m \frac{B_{up}}{M} \log_2(1+\Upsilon_{n,i}^m)} + \frac{m_i}{\beta_{j,i}^k \frac{B_{down}}{M} \log_2(1+\Psi_{j,i}^k)} \right) + \sum_{i=1}^{U} \omega_E^i \left( \sum_{\delta=1}^{s_i} \xi_i c_i^2 \varphi_i f_{l_\delta} + \sum_{\delta=s_i+1}^{\mathcal{F}} \xi_e (\lambda(r_i)c_{min})^2 \varphi_e f_{l_\delta} + p_{n,i}^m \cdot \frac{w_{s_i}}{\beta_{n,i}^m \cdot \frac{B_{up}}{M} \log_2(1+\Upsilon_{n,i}^m)} + P_{j,i}^k \cdot \frac{m_i}{\beta_{j,i}^k \frac{B_{down}}{M} \log_2(1+\Psi_{j,i}^k)} \right) \quad (21)$$

where $R_{n,i}^m$ and $\Phi_{j,i}^k$ are presented in (6) and (9), respectively.

In (21), the size of tasks that calculated on mobile device and edge server relates to $s_i$, i.e., $\sum_{\delta=1}^{s_i} f_{l_\delta}$ and $\sum_{\delta=s_i+1}^{\mathcal{F}} f_{e_\delta}$, and $w_{s_i}$ relates to the intermediate data that transmitted between end device and edge server, which cannot be relaxed as continuous variables. Therefore, we define two variables as follows. For user $i$, let $f_l^{i-\delta} = \sum_{\delta=1}^{s_i} f_{l_\delta}$ and $s_i \in \{1,2,...,\mathcal{F}\}$, then $f_l^{i-1} = f_{l_1}$, $f_l^{i-2} = f_{l_1} + f_{l_2}$, and so on. Then we can change variable $s_i$ to $f_l^i$, and $f_l^i \in \{f_l^{i-1}, f_l^{i-2}, ..., f_l^{i-\mathcal{F}}\}$, where $f_{l_\delta}$ is calculated based on (2). Let $Z_i = \sum_{\delta=1}^{\mathcal{F}} f_{l_\delta}$ is the size of all layers, then $f_e^i = Z_i - f_l^i$. Moreover, $f_l^i$, $f_e^i$, and $w_{s_i}$ are calculated by mobile users in advance and stored in devices with inference model.

Then, we introduce $f_l^i$, $f_e^i$, and $w_{s_i}$ into (18). For different layers of model, we can get a series of utility functions $\Gamma = \{\Gamma_1, \Gamma_2, ..., \Gamma_{s_i}, ..., \Gamma_{\mathcal{F}}\}$, where $\Gamma_{s_i}$ can be expressed as:

$$\Gamma_{s_i} = \sum_{i=1}^{U} \omega_T^i \left( \frac{f_l^i}{c_i} + \frac{f_e^i}{\lambda(r_i)c_{min}} + \frac{w_{s_i}}{R_{n,i}^m} + \frac{m_i}{\Phi_{j,i}^k} \right)$$
$$+ \sum_{i=1}^{U} \omega_E^i \left( \xi_i c_i^2 \varphi_i f_l^i + \xi_e(\lambda(r_i)c_{min})^2 \varphi_e f_e^i + p_{n,i}^m \cdot \frac{w_{s_i}}{R_{n,i}^m} + P_{j,i}^k \cdot \frac{m_i}{\Phi_{j,i}^k} \right) \quad (22)$$

where $f_l^i$, $f_e^i$, and $w_{s_i}$ are already known in advance for each inference model in mobile device.

In (22), there are $U$ mobile users. For each mobile user, we need to calculate the optimal $B$, $P$, and $r$. Since the parameters $P$ and $r$ are continuous, the variable spaces of $P$ and $r$ are large and infinite dimensional. Additionally, since $B$, $P$, and $r$ are all related to $s_i$, and $B$ and $P$ are close coupled, it is difficult to calculate the optimal value of $B$, $P$, and $r$ separately. Thus, in this paper, for finding the optimal tradeoff between inference delay and energy consumption, we introduce the gradient descent approach into our algorithm. For using the gradient descent to address above issue, firstly, we need to prove that the weight function shown in (21) is differentiable which is defined as follows.

**Definition 1 [43].** For $f(x,y,z)$, if its partial derivative on $x$, $y$, and $z$, i.e., $f'(x,y,z)|_x$, $f'(x,y,z)|_y$, and $f'(x,y,z)|_z$, exist and continue, the $f(x,y,z)$ is differentiable.

Based on Definition 1, we have the conclusion as follows.

**Corollary 1.** When the values of $f_l^i$, $f_e^i$, and $w_{s_i}$ are know in advance, and we loose the constraints of $\beta_{n,i}^m \in \{0,1\}$ and $\beta_{j,i}^k \in \{0,1\}$ to $\beta_{n,i}^m \in [0\ 1]$ and $\beta_{j,i}^k \in [0\ 1]$, the utility function shown in (21) is differentiable.

*Proof.* Since the values of $f_l^i$, $f_e^i$, and $w_{s_i}$ are already know in advance for every inference model in each mobile device, the values of $y_1 = \frac{f_l^i}{c_i}$ and $y_2 = \xi_i c_i^2 \varphi_i f_l^i$ can be calculated easily and nothing to do with the partial derivative on $B_i$, $P_i$, and $r_i$. Therefore, the partial derivative of $\Gamma_{s_i}$ on $B_i$, $P_i$, and $r_i$ can be expressed as:

$$\Gamma'_{s_i}|_{\beta_{n,i}^m} = \sum_{i=1}^{U} \left( \omega_T^i \cdot \frac{w_{s_i}}{R_{n,i}^m} \right)\bigg|_{\beta_{n,i}^m} + \sum_{i=1}^{U} \left( \omega_E^i \cdot p_{n,i}^m \cdot \frac{w_{s_i}}{R_{n,i}^m} \right)\bigg|_{\beta_{n,i}^m} \quad (23)$$

$$\Gamma'_{s_i}|_{\beta_{j,i}^k} = \sum_{i=1}^{U} \left( \omega_T^i \cdot \frac{m_i}{\Phi_{j,i}^k} \right)\bigg|_{\beta_{j,i}^k} + \sum_{i=1}^{U} \left( \omega_E^i \cdot P_{j,i}^k \cdot \frac{m_i}{\Phi_{j,i}^k} \right)\bigg|_{\beta_{j,i}^k} \quad (24)$$

$$\Gamma'_{s_i}|_{p_{n,i}^m} = \sum_{i=1}^{U} \left( \omega_T^i \cdot \frac{w_{s_i}}{R_{n,i}^m} \right)\bigg|_{p_{n,i}^m} + \sum_{i=1}^{U} \left( \omega_E^i \cdot p_{n,i}^m \cdot \frac{w_{s_i}}{R_{n,i}^m} \right)\bigg|_{p_{n,i}^m} \quad (25)$$

$$\Gamma'_{s_i}|_{P_{j,i}^k} = \sum_{i=1}^{U} \left( \omega_T^i \cdot \frac{m_i}{\Phi_{j,i}^k} \right)\bigg|_{P_{j,i}^k} + \sum_{i=1}^{U} \left( \omega_E^i \cdot P_{j,i}^k \cdot \frac{m_i}{\Phi_{j,i}^k} \right)\bigg|_{P_{j,i}^k} \quad (26)$$

$$\Gamma'_{s_i}|_{r_i} = \sum_{i=1}^{U} \left( \omega_T^i \cdot \frac{f_e^i}{\lambda(r_i)c_{min}} \right)\bigg|_{r_i} + \sum_{i=1}^{U} \left( \omega_E^i \cdot \xi_e(\lambda(r_i)c_{min})^2 \varphi_e f_e^i \right)\bigg|_{r_i} \quad (27)$$

Moreover, according to (23), let $\Delta = \sum_{l=1, l\neq n}^{N} \sum_{t=1}^{U} \beta_{l,t}^m p_{l,t}^m |g_{l,t}^m|^2 + \sigma^2$, we have:

$$\sum_{i=1}^{U} \left( \omega_T^i \cdot \frac{w_{s_i}}{R_{n,i}^m} \right)\bigg|_{\beta_{n,i}^m} = \omega_T^1 \frac{w_{s_i}}{R_{n,1}^m}\bigg|_{\beta_{n,i}^m} + \omega_T^2 \frac{w_{s_i}}{R_{n,2}^m}\bigg|_{\beta_{n,i}^m} + \cdots + \omega_T^i \frac{w_{s_i}}{R_{n,i}^m}\bigg|_{\beta_{n,i}^m} + \cdots + \omega_T^U \frac{w_{s_i}}{R_{n,U}^m}\bigg|_{\beta_{n,i}^m} \quad (28)$$

$$\omega_T^i \frac{w_{s_i}}{R_{n,i}^m}\bigg|_{\beta_{n,i}^m} = -\omega_T^i w_{s_i} \frac{\frac{B_{up}}{M} \log_2\left(1+\frac{p_{n,i}^m |h_{n,i}^m|^2}{\sum_{v=1,v\neq i}^{N} \beta_{n,v}^m p_{n,v}^m |h_{n,v}^m|^2 + \Delta}\right)}{\left\{\beta_{n,i}^m \frac{B_{up}}{M} \log_2\left(1+\frac{p_{n,i}^m |h_{n,i}^m|^2}{\sum_{v=1,v\neq i}^{N} \beta_{n,v}^m p_{n,v}^m |h_{n,v}^m|^2 + \Delta}\right)\right\}^2}$$
$$= -\omega_T^1 w_{s_i} \frac{1}{(\beta_{n,i}^m)^2 \left\{\frac{B_{up}}{M} \log_2\left(1+\frac{p_{n,i}^m |h_{n,i}^m|^2}{\sum_{v=1,v\neq i}^{N} \beta_{n,v}^m p_{n,v}^m |h_{n,v}^m|^2 + \Delta}\right)\right\}^2} \quad (29)$$

$$\sum_{\tau=1,\tau\neq i}^{U} \left( \omega_T^\tau \cdot \frac{w_{s_i}}{R_{n,\tau}^m} \right)\bigg|_{\beta_{n,i}^m} = \omega_T^1 \frac{w_{s_i}}{R_{n,1}^m}\bigg|_{\beta_{n,1}^m} + \cdots + \omega_T^{i-1} \frac{w_{s_i}}{R_{n,i-1}^m}\bigg|_{\beta_{n,i-1}^m} + \omega_T^{i+1} \frac{w_{s_i}}{R_{n,i+1}^m}\bigg|_{\beta_{n,i+1}^m} + \cdots + \omega_T^U \frac{w_{s_i}}{R_{n,U}^m}\bigg|_{\beta_{n,U}^m}$$
$$= \sum_{\tau=1,\tau\neq i}^{U} \frac{\beta_{n,\tau}^m \frac{B_{up}}{M}}{\left(1+\frac{p_{n,\tau}^m |h_{n,\tau}^m|^2}{\sum_{v=1,v\neq \tau}^{N} \beta_{n,v}^m p_{n,v}^m |h_{n,v}^m|^2 + \Delta}\right) \ln 2} \cdot \frac{p_{n,\tau}^m |h_{n,\tau}^m|^2 \cdot p_{n,i}^m |h_{n,i}^m|^2}{\left(\sum_{v=1,v\neq \tau}^{N} \beta_{n,v}^m p_{n,v}^m |h_{n,v}^m|^2 + \Delta\right)^2} \quad (30)$$

The calculation of $\sum_{i=1}^{U} \left( \omega_E^i \cdot p_{n,i}^m \cdot \frac{w_{s_i}}{R_{n,i}^m} \right)\bigg|_{\beta_{n,i}^m}$, $\Gamma'_{s_i}|_{\beta_{j,i}^k}$, $\Gamma'_{s_i}|_{p_{n,i}^m}$, and $\Gamma'_{s_i}|_{P_{j,i}^k}$ are similar with that shown in (29). For $\Gamma'_{s_i}|_{\beta_{n,i}^m}$ shown in (23) and $\forall \beta_{n,i}^m \in [0\ 1]$, we have $x_1 =$

$-\frac{\omega_T^1 w_{s_i}}{(\beta_{n,i}^m)^2 \left\{\frac{B_{up}}{M}\log_2\left(1+\frac{p_{n,i}^m|h_{n,i}^m|^2}{\sum_{v=1,v\neq i}^N \beta_{n,v}^m p_{n,v}^m |h_{n,v}^m|^2+\Delta}\right)\right\}^2}$ is continuous, $x_2 =$

$\sum_{\tau=1,\tau\neq i}^U \frac{\beta_{n,\tau}^m \frac{B_{up}}{M}}{\left(1+\frac{p_{n,\tau}^m|h_{n,\tau}^m|^2}{\sum_{v=1,v\neq \tau}^N \beta_{n,v}^m p_{n,v}^m|h_{n,v}^m|^2+\Delta}\right)\ln 2} \cdot \frac{p_{n,\tau}^m|h_{n,\tau}^m|^2 \cdot p_{n,i}^m|h_{n,i}^m|^2}{\left(\sum_{v=1,v\neq \tau}^N \beta_{n,v}^m p_{n,v}^m|h_{n,v}^m|^2+\Delta\right)^2}$ is

continuous, and $x_3 = \sum_{i=1}^U \left(\omega_E^i \cdot p_{n,i}^m \cdot \frac{w_{s_i}}{R_{n,i}^m}\right)\Big|_{\beta_{n,i}^m}$ is continuous. Thus, based on the operational rule of continuous function, the $\Gamma'_{s_i}|_{\beta_{n,i}^m}$ is continuous with $\forall \beta_{n,i}^m \in [0\ 1]$. Similarly, the $\Gamma'_{s_i}|_{\beta_{j,i}^k}$, $\Gamma'_{s_i}|_{p_{n,i}^m}$, and $\Gamma'_{s_i}|_{P_{j,i}^k}$ are all continuous. Additionally, since $\lambda(r_i)$ is continuous, $\Gamma'_{s_i}|_{r_i}$ is continuous with $\forall r_i \in [r_{min}\ r_{max}]$. Therefore, Corollary 1 is proved. ∎

The Corollary 1 means that when the values of $f_l^i$, $f_e^i$, and $w_{s_i}$ are known, the GD approach can be used in (22) to find the optimal strategies of $B$, $P$, and $r$. However, the utility function shown in (22) is only the utility when the model segmentation point is $s_i$, there are $\mathcal{F}$ layers in the inference model, which means that the GD algorithm needs to be repeated $\mathcal{F}$ times to find the global optimal solutions for $B$, $P$, and $r$. However, considering the complexity and convergence time of GD approach, repeating the GD approach $\mathcal{F}$ times will cause serious delay and complexity. Fortunately, for the GD approach, if we can select the initial value carefully, the complexity and convergence time can be reduced greatly. Therefore, in this paper, based on the greedy approach, we propose the Loop iteration GD algorithm, shorted as Li-GD. The details of the Li-GD algorithm are presented in TABLE I.

TABLE I

| Algorithm 1: Loop iteration GD algorithm (Li-GD) |
|---|
| **Input:** |
| Objective function: $\boldsymbol{\Gamma} = \{\Gamma_1, \Gamma_2, \dots, \Gamma_{s_i}, \dots, \Gamma_{\mathcal{F}}\}$; |
| Gradient function: $\boldsymbol{\nabla} = \{\nabla_{B_i} = \frac{\partial \Gamma_{s_i}}{\partial B_i}, \nabla_{P_i} = \frac{\partial \Gamma_{s_i}}{\partial P_i}, \nabla_{r_i} = \frac{\partial \Gamma_{s_i}}{\partial r_i}\}$; |
| Algorithm accuracy: $\varepsilon$; |
| Step size: $\lambda$; |
| **Output:** |
| The optimal solution $\boldsymbol{O}^* = \{\boldsymbol{B}^*, \boldsymbol{P}^*, \boldsymbol{r}^*\}$; |
| 1. Let $\boldsymbol{B}^{j(0)} \in [0\ 1]$, $P^{j(0)} \in [P_{min}\ P_{max}]$, and $\boldsymbol{r}^{j(0)} \in [r_{min}\ r_{max}]$, $\forall i \in [1\ U]$ and $\forall j \in [1\ \mathcal{F}]$; |
| *# Calculating the optimal strategy for the first layer #* |
| 2. If $j = 1$; |
| 3. Let $k \leftarrow 0$, $\boldsymbol{B}^{j(k)} = \{B_1^{j(k)}, \dots, B_{\mathcal{F}}^{j(k)}\}$, $\boldsymbol{P}^{j(k)} = \{P_1^{j(k)}, \dots, P_{\mathcal{F}}^{j(k)}\}$ and $\boldsymbol{r}^{j(k)} = \{r_1^{j(k)}, \dots, r_{\mathcal{F}}^{j(k)}\}$; |
| 4. Calculating $\Gamma_{s_i}(\boldsymbol{B}^{j(k)}, \boldsymbol{P}^{j(k)}, \boldsymbol{r}^{j(k)})$; |
| 5. Calculating the gradient $\boldsymbol{g}_k = g(\boldsymbol{B}^{j(k)}, \boldsymbol{P}^{j(k)}, \boldsymbol{r}^{j(k)})$; |
| 6. If $\|g_k\| < \varepsilon$, then $\boldsymbol{B}^{j*} \leftarrow \boldsymbol{B}^{j(k)}$, $\boldsymbol{P}^{j*} \leftarrow \boldsymbol{P}^{j(k)}$ and $\boldsymbol{r}^{j*} \leftarrow \boldsymbol{r}^{j(k)}$; |
| 7. Otherwise, let $\boldsymbol{\zeta}_k = -g(\boldsymbol{B}^{j(k)}, \boldsymbol{P}^{j(k)}, \boldsymbol{r}^{j(k)})$, and let $\boldsymbol{B}^{j(k+1)} = \boldsymbol{B}^{j(k)} + \lambda \boldsymbol{\zeta}_k$, $\boldsymbol{P}^{j(k+1)} = \boldsymbol{P}^{j(k)} + \lambda \boldsymbol{\zeta}_k$, and $\boldsymbol{r}^{j(k+1)} = \boldsymbol{r}^{j(k)} + \lambda \boldsymbol{\zeta}_k$; |
| 8. Calculating $\Gamma_{s_i}(\boldsymbol{B}^{j(k+1)}, \boldsymbol{P}^{j(k+1)}, \boldsymbol{r}^{j(k+1)}) = \Gamma_{s_i}(\boldsymbol{B}^{j(k)} + \lambda \boldsymbol{p}_k, \boldsymbol{P}^{j(k)} + \lambda \boldsymbol{\zeta}_k, \boldsymbol{r}^{j(k)} + \lambda \boldsymbol{p}_k)$; |
| 9. If $\|\Gamma_{s_i}(\boldsymbol{B}^{j(k+1)}, \boldsymbol{P}^{j(k+1)}, \boldsymbol{r}^{j(k+1)}) - \Gamma_{s_i}(\boldsymbol{B}^{j(k)}, \boldsymbol{P}^{j(k)}, \boldsymbol{r}^{j(k)})\| < \varepsilon$ or $\max\{\|\boldsymbol{B}^{j(k+1)} - \boldsymbol{B}^{j(k)}\|, \|\boldsymbol{P}^{j(k+1)} - \boldsymbol{P}^{j(k)}\|, \|\boldsymbol{r}^{j(k+1)} - \boldsymbol{r}^{j(k)}\|\} < \varepsilon$; |
| 10. then $\boldsymbol{B}^{j*} \leftarrow \boldsymbol{B}^{j(k+1)}$, $\boldsymbol{P}^{j*} \leftarrow \boldsymbol{P}^{j(k+1)}$, and $\boldsymbol{r}^{j*} \leftarrow \boldsymbol{r}^{j(k+1)}$; |
| 11. otherwise, $k = k+1$; |
| 12. end if |
| *# Calculating the optimal strategy of the rest layers #* |
| 13. When $1 < j \leq \mathcal{F}$; |
| *# Loop iteration #* |
| 14. Let $\boldsymbol{B}^{j+1(0)} = \boldsymbol{B}^{j*}$, $\boldsymbol{P}^{j+1(0)} = \boldsymbol{P}^{j*}$, and $\boldsymbol{r}^{j+1(0)} = \boldsymbol{r}^{j*}$, $\forall i \in [1\ U]$ and $\forall j \in [1\ \mathcal{F}]$; |
| 15. repeating step 3 to Step 11; |
| 16. $j = j+1$; |
| *# Finding the optimal strategy #* |
| 17. Calculating $\boldsymbol{\Gamma} = \{\Gamma_1(\boldsymbol{B}^{1*}, \boldsymbol{P}^{1*}, \boldsymbol{r}^{1*}), \dots, \Gamma_{\mathcal{F}}(\boldsymbol{B}^{\mathcal{F}*}, \boldsymbol{P}^{\mathcal{F}*}, \boldsymbol{r}^{\mathcal{F}*})\}$; |
| 18. $(\boldsymbol{s}, \boldsymbol{B}, \boldsymbol{P}, \boldsymbol{r}) \leftarrow \arg\min_{s^*, \boldsymbol{B}^*, \boldsymbol{P}^*, \boldsymbol{r}^*} \boldsymbol{\Gamma}$; |
| 19. If $B > 0.5 \rightarrow B = 1$; |
| 20. otherwise $B = 0$. |

The Li-GD algorithm presented in TABLE I is composed of three parts.

1) **(Line2-Line12):** Calculating the optimal strategy when the model segmentation point is in the first layer. For the Li-GD algorithm, since the model segmentation strategy is discrete, we need to calculate the optimal resource allocation strategies layer by layer. For the first layer, its starting values are $\boldsymbol{B}^{1(0)} = \{B_1^{1(0)}, \dots, B_U^{1(0)}\}$, $\boldsymbol{P}^{1(0)} = \{P_1^{1(0)}, \dots, P_U^{1(0)}\}$ and $\boldsymbol{r}^{1(0)} = \{r_1^{1(0)}, \dots, r_U^{1(0)}\}$, where $\boldsymbol{B}^{1(0)} \in [0\ 1]$, $\boldsymbol{r}^{1(0)} \in [r_{min}\ r_{max}]$, and $\boldsymbol{P}^{1(0)} \in [p_{min}\ p_{max}]$. Additionally, they are selected without any information of the final optimal values. Then, the GD algorithm is executed with step size $\lambda$ and gradient $-\boldsymbol{g}_k$. After $k$ rounds iteration, when the threshold of accuracy is reached, the optimal solutions of resource allocation strategy for the first layer are $\boldsymbol{B}^{1*} \leftarrow \boldsymbol{B}^{1(k)}$, $\boldsymbol{P}^{1*} \leftarrow \boldsymbol{P}^{1(k)}$, and $\boldsymbol{r}^{1*} \leftarrow \boldsymbol{r}^{1(k)}$.

2) **(Line13-Line16):** Calculating the optimal resource strategy for the rest layers. When the optimal resource allocation strategy of the first layer is calculated, then from the second layer, the starting values of this layer are the optimal values of the last layer. For instance, $\boldsymbol{B}^{2(0)} = \boldsymbol{B}^{1*}$, $\boldsymbol{P}^{2(0)} = \boldsymbol{P}^{1*}$ and $\boldsymbol{r}^{2(0)} = \boldsymbol{r}^{1*}$, $\boldsymbol{B}^{3(0)} = \boldsymbol{B}^{2*}$, $\boldsymbol{P}^{3(0)} = \boldsymbol{P}^{2*}$, and $\boldsymbol{r}^{3(0)} = \boldsymbol{r}^{2*}$, etc. The GD process is the same as that when calculating the optimal strategy for first layer. In this stage, the optimal resource allocation strategies for $2th$ layer to the $\mathcal{F}th$ layer are calculated, which are $\boldsymbol{B}^{(2-\mathcal{F})*} = \{\boldsymbol{B}^{2*}, \dots, \boldsymbol{B}^{\mathcal{F}*}\}$, $\boldsymbol{P}^{(2-\mathcal{F})*} = \{\boldsymbol{P}^{2*}, \dots, \boldsymbol{P}^{\mathcal{F}*}\}$, and $\boldsymbol{r}^{(2-\mathcal{F})*} = \{\boldsymbol{r}^{2*}, \dots, \boldsymbol{r}^{\mathcal{F}*}\}$.

3) **(Line17-Line20)**: Finding the final optimal model segmentation strategy and resource allocation strategy. When the optimal resource allocation strategies for all the layers are calculated, which are $\boldsymbol{B}^* = \{\boldsymbol{B}^{1*}, \boldsymbol{B}^{2*}, \dots, \boldsymbol{B}^{\mathcal{F}*}\}$, $\boldsymbol{P}^* = \{\boldsymbol{P}^{1*}, \boldsymbol{P}^{2*}, \dots, \boldsymbol{P}^{\mathcal{F}*}\}$, and $\boldsymbol{r}^* = \{\boldsymbol{r}^{1*}, \boldsymbol{r}^{2*}, \dots, \boldsymbol{r}^{\mathcal{F}*}\}$, respectively, then substituting the $\boldsymbol{B}^*$, $\boldsymbol{P}^*$, and $\boldsymbol{r}^*$ into (18) and getting $\mathcal{F}$ utility values $\boldsymbol{U}^* = \{U^{1*}, U^{2*}, \dots, U^{\mathcal{F}*}\}$. Finally, finding the minimum value from $\boldsymbol{U}^*$, and the model split strategy and resource allocation strategy that associated with this utility value is selected as the final optimal strategy. Finally, since the value range of $\boldsymbol{B}$ is changed from $\{0,1\}$ to $[0\ 1]$, we give the approximate rule as: if $B > 0.5$, $B = 1$; otherwise $B = 0$.

The theory foundations of Li-GD approach are: 1) for the GD algorithm, carefully selecting the start value can decrease the complexity and speed up the convergence greatly [51]; 2) for the inference model, in most cases, the optimal resource allocation strategy between adjacent layers is much more similarly than the nonadjacent layers. Since the size of layers and the intermediate transmission data between adjacent layers are similar [9]. For proving the effectiveness of the Li-GD algorithm that proposed in this section, we give the conclusions as follows.

### B. The properties of Li-GD algorithm

In this section, we investigate the properties of Li-GD algorithm, including convergence, complexity, and approximate error. The details are shown below.

**Corollary 2.** The Li-GD algorithm is convergent, and the convergence time is $K = \frac{\|x^0-x^*\|_2^2}{2\eta\epsilon}$, where $\eta$ is the step size and $\eta \leq \frac{1}{L}$, $\epsilon$ is the threshold of accuracy.

*Proof.* Based on the conclusions in [51], if the differentiable function $f(x)$ satisfies: 1) L-Lipschitz smooth and 2) convex, the $f(x)$ is convergent.

Based on (22), let $\nabla = \sum_{x=1,x\neq j}^{N}\sum_{y=1}^{U}\beta_{x,y}^k P_{x,y}^k |G_{x,y}^k|^2 + \sigma^2$, then the parts of $\Gamma_{S_i}$ relate to $B_i$ are:

$$\Gamma_{S_i}(B_i) = \sum_{i=1}^{U}\omega_T^i\left(\frac{w_{S_i}}{R_{n,i}^m} + \frac{m_i}{\Phi_{j,i}^k}\right) + \sum_{i=1}^{U}\omega_E^i\left(p_{n,i}^m \cdot \frac{w_{S_i}}{R_{n,i}^m} + P_{j,i}^k \cdot \frac{m_i}{\Phi_{j,i}^k}\right)$$

$$= \sum_{i=1}^{U}\omega_T^i\left(\frac{w_{S_i}}{\beta_{n,i}^m \frac{B_{up}}{M}\log_2\left(1+\frac{p_{n,i}^m|h_{n,i}^m|^2}{\sum_{v=u+1}^{U}\beta_{n,v}^m p_{n,v}^m|h_{n,v}^m|^2+\Delta}\right)}\right.$$

$$\left. + \frac{m_i}{\beta_{j,i}^k \frac{B_{down}}{M}\log_2\left(1+\frac{|H_{j,i}^k|^2 P_{j,i}^k}{\sum_{q=u+1}^{U}\beta_{j,q}^k P_{j,q}^k|H_{j,q}^k|^2+\nabla}\right)}\right) + \sum_{i=1}^{U}\omega_E^i\left(p_{n,i}^m \cdot \right.$$

$$\frac{w_{S_i}}{\beta_{n,i}^m \frac{B_{up}}{M}\log_2\left(1+\frac{p_{n,i}^m|h_{n,i}^m|^2}{\sum_{v=u+1}^{U}\beta_{n,v}^m p_{n,v}^m|h_{n,v}^m|^2+\Delta}\right)} + P_{j,i}^k \cdot$$

$$\left. \frac{m_i}{\beta_{j,i}^k \frac{B_{down}}{M}\log_2\left(1+\frac{|H_{j,i}^k|^2 P_{j,i}^k}{\sum_{q=u+1}^{U}\beta_{j,q}^k P_{j,q}^k|H_{j,q}^k|^2+\nabla}\right)}\right) \quad (31)$$

Let:

$$A = \frac{w_{S_i}}{\beta_{n,i}^m \frac{B_{up}}{M}\log_2\left(1+\frac{p_{n,i}^m|h_{n,i}^m|^2}{\sum_{v=u+1}^{U}\beta_{n,v}^m p_{n,v}^m|h_{n,v}^m|^2+\Delta}\right)} \quad (32)$$

$$B = \frac{m_i}{\beta_{j,i}^k \frac{B_{down}}{M}\log_2\left(1+\frac{|H_{j,i}^k|^2 P_{j,i}^k}{\sum_{q=u+1}^{U}\beta_{j,q}^k P_{j,q}^k|H_{j,q}^k|^2+\nabla}\right)} \quad (33)$$

For $\Gamma_{S_i}(B_i)$, $A$ and $B$ are similar for uplink subchannel allocation $\beta_{n,i}^m$ and downlink subchannel allocation $\beta_{j,i}^k$, which can be simplified as:

$$f(x) = \frac{1}{x \log_2\left(1+\frac{1}{x}\right)} \quad (34)$$

And the first-order derivative of (34) can be expressed as:

$$f'(x) = \frac{1}{x^2 \log_2\left(1+\frac{1}{x}\right)}\left(\frac{1}{(1+x)\ln 2 \log_2\left(1+\frac{1}{x}\right)} - 1\right) \quad (35)$$

*L-Lipschitz smooth:*

Let $y = kx$ and $k > 1$, then we have $|x - y| = x|k - 1|$ and:

$$|f'(x) - f'(y)| = \left|\frac{1}{k^2 x^2 \log_2\left(1+\frac{1}{kx}\right)}\left(\frac{1}{(1+kx)\ln 2 \log_2\left(1+\frac{1}{kx}\right)} - 1\right) - \frac{1}{x^2 \log_2\left(1+\frac{1}{x}\right)}\left(\frac{1}{(1+x)\ln 2 \log_2\left(1+\frac{1}{x}\right)} - 1\right)\right| \quad (36)$$

Since $\frac{1}{\log_2\left(1+\frac{1}{x}\right)}$ increases with the increasing of $x$ and $\frac{1}{kx}$ decreases with the increasing of $x$, we have:

$$\frac{1}{k^2 x^2 \log_2\left(1+\frac{1}{kx}\right)}\left(\frac{1}{(1+kx)\ln 2 \log_2\left(1+\frac{1}{kx}\right)} - 1\right)$$

$$< \frac{1}{kx^2 \log_2\left(1+\frac{1}{x}\right)}\left(\frac{1}{(1+x)\ln 2 \log_2\left(1+\frac{1}{kx}\right)} - 1\right) \quad (37)$$

Firstly, let:

$$\varphi = \frac{\frac{1}{kx^2 \log_2\left(1+\frac{1}{x}\right)}}{\frac{1}{k^2 x^2 \log_2\left(1+\frac{1}{kx}\right)}} = \frac{k^2 x^2 \log_2\left(1+\frac{1}{kx}\right)}{kx^2 \log_2\left(1+\frac{1}{x}\right)} = \frac{k \log_2\left(1+\frac{1}{kx}\right)}{\log_2\left(1+\frac{1}{x}\right)} \quad (38)$$

Since $0 < x \leq 1$, we have: $\log_2\left(1+\frac{1}{x}\right) > 1$ and $\log_2\left(1+\frac{1}{kx}\right) > 1$. Therefore:

$$\theta = \frac{k 2^{\log_2\left(1+\frac{1}{kx}\right)}}{2^{\log_2\left(1+\frac{1}{x}\right)}} = \frac{kx+1}{1+x} > 1 \quad (39)$$

Thus, (39) means $k \log_2\left(1+\frac{1}{kx}\right) > \log_2\left(1+\frac{1}{x}\right)$, which also means $\varphi > 1$. Moreover, it is obvious that $\frac{1}{(1+kx)\ln 2 \log_2\left(1+\frac{1}{kx}\right)} - 1 < \frac{1}{(1+x)\ln 2 \log_2\left(1+\frac{1}{x}\right)} - 1$.

Therefore, (36) can be rewritten as:

$$|f'(x) - f'(y)| \leq \left|\frac{1}{kx^2 \log_2\left(1+\frac{1}{x}\right)}\left(\frac{1}{(1+x)\ln 2 \log_2\left(1+\frac{1}{kx}\right)} - 1\right) - \frac{1}{x^2 \log_2\left(1+\frac{1}{x}\right)}\left(\frac{1}{(1+x)\ln 2 \log_2\left(1+\frac{1}{x}\right)} - 1\right)\right|$$

$$= \frac{1}{x^2(1+x)\ln 2 \log_2\left(1+\frac{1}{x}\right)}\left|\frac{1}{k \log_2\left(1+\frac{1}{kx}\right)} - \frac{1}{\log_2\left(1+\frac{1}{x}\right)}\right|$$

$$\leq \frac{1}{x^2(1+x)\ln 2 \log_2\left(1+\frac{1}{x}\right)}\left|\frac{1}{\log_2\left(1+\frac{1}{kx}\right)} - \frac{1}{\log_2\left(1+\frac{1}{x}\right)}\right| \quad (40)$$

If the (40) is no larger than $L|x - y|$ holds, then the following constraints should be satisfied:

$$\frac{1}{x^2(1+x)\ln 2 \log_2\left(1+\frac{1}{x}\right)}\left|\frac{1}{\log_2\left(1+\frac{1}{kx}\right)} - \frac{1}{\log_2\left(1+\frac{1}{x}\right)}\right|$$
$$< Lx|k-1| = L|x - y| \quad (41)$$

The (41) equals to:

$$L > \frac{1}{(k-1)x^3(1+x)\ln 2 \log_2\left(1+\frac{1}{x}\right)}\left|\frac{1}{\log_2\left(1+\frac{1}{kx}\right)} - \frac{1}{\log_2\left(1+\frac{1}{x}\right)}\right| \quad (42)$$

Moreover, it is easy to be proved that $j(x) = \frac{1}{\log_2\left(1+\frac{1}{kx}\right)} - \frac{1}{\log_2\left(1+\frac{1}{x}\right)}$ is monotone increasing with $x$, which is presented as follows.

$$j'(x) = \frac{1}{kx^2\left(1+\frac{1}{kx}\right)\ln 2 \log_2\left(1+\frac{1}{kx}\right)} - \frac{1}{x^2\left(1+\frac{1}{x}\right)\ln 2 \log_2\left(1+\frac{1}{x}\right)} \quad (43)$$

Let:

$$l(x) = \frac{\frac{1}{kx^2\left(1+\frac{1}{kx}\right)\ln 2\left(\log_2\left(1+\frac{1}{kx}\right)\right)^2}}{\frac{1}{x^2\left(1+\frac{1}{x}\right)\ln 2\left(\log_2\left(1+\frac{1}{x}\right)\right)^2}} = \frac{x^2\left(1+\frac{1}{x}\right)\ln 2\left(\log_2\left(1+\frac{1}{x}\right)\right)^2}{kx^2\left(1+\frac{1}{kx}\right)\ln 2\left(\log_2\left(1+\frac{1}{kx}\right)\right)^2}$$

$$= \frac{(1+x)\left(\log_2\left(1+\frac{1}{x}\right)\right)^2}{(1+kx)\left(\log_2\left(1+\frac{1}{kx}\right)\right)^2} > \frac{\left(\log_2\left(1+\frac{1}{x}\right)\right)^2}{k\left(\log_2\left(1+\frac{1}{kx}\right)\right)^2} > \frac{\left(\log_2\left(\frac{1}{k}+\frac{1}{x}\right)\right)^2}{k\left(\log_2\left(1+\frac{1}{kx}\right)\right)^2}$$

$$= \frac{(\log_2(k+x) - \log_2(kx))^2}{k(\log_2(1+kx) - \log_2(kx))^2} = \left(\frac{\log_2(k+x)}{\log_2(1+kx)^{\sqrt{k}}}\right)^2 \quad (44)$$

Let $\eta = \sqrt{k} > 1$, then we have:

$$v(x) = k + x - (1+kx)^\eta \quad (45)$$

According to (45), we have:

$$v'(x) = 1 - \eta k(1+kx)^{\eta-1} < 0 \quad (46)$$

Therefore, for $v(x)$, when $x = 0$, it has the maximum value $v(0) = k - 1 > 0$. Thus, we can conclude that $\frac{\log_2(k+x)}{\log_2(1+kx)^{\sqrt{k}}} > 1$, which means that $l(x) > 1$ and $j'(x) > 0$.

Thus, for (42), we can conclude that:
$$L > \frac{1}{2(k-1)\ln 2}\left|\frac{1}{\log_2\left(1+\frac{1}{k}\right)} - 1\right| \quad (47)$$

Since $k > 1$, $\frac{1}{2(k-1)\ln 2}\left|\frac{1}{\log_2\left(1+\frac{1}{k}\right)} - 1\right| < 1$. Therefore, according to (47), there must exist $L > 1$ can satisfy the constraint in (41).

According to the same method, the $\Gamma_{s_i}$ relates to $P_i$ is the same as (31). Therefore, the $\Gamma_{s_i}(P_i)$ can be simplified to four basic functions $y(x) = \frac{1}{\log_2\left(1+\frac{1}{x}\right)}$, $g(x) = \frac{1}{\log_2(1+x)}$, $h(x) = \frac{x}{\log_2(1+x)}$, and $z(x) = \frac{x}{\log_2\left(1+\frac{1}{x}\right)}$. The same as the process that shown above, the $y(x)$, $g(x)$, $h(x)$, and $z(x)$ are all L-Lipschitz smooth.

*Convex:*

For the differentiable function $f(x)$, if $f''(x) > 0$, it is convex [51]. The $f'(x)$ is presented in (35). Thus, let $\mathcal{L} = \frac{1}{x^2 \log_2\left(1+\frac{1}{x}\right)}$, $\mathcal{M} = \frac{1}{(1+x)\ln 2 \log_2\left(1+\frac{1}{x}\right)} - 1$, then we have:

$$\mathcal{L}' = \frac{1}{x^3 \log_2\left(1+\frac{1}{x}\right)}\left[\frac{1}{(1+x)\ln 2 \log_2\left(1+\frac{1}{x}\right)} - 2\right] \quad (48)$$

$$\mathcal{M}' = \frac{1}{(1+x)^3 \ln 2 \log_2\left(1+\frac{1}{x}\right)}\left[\frac{1}{x \ln 2 \log_2\left(1+\frac{1}{x}\right)} - 1\right] \quad (49)$$

Therefore,
$$f''(x) = \frac{1}{x^3 \log_2\left(1+\frac{1}{x}\right)}\left[\frac{1}{(1+x)\ln 2 \log_2\left(1+\frac{1}{x}\right)} - 2\right] \cdot \left[\frac{1}{(1+x)\ln 2 \log_2\left(1+\frac{1}{x}\right)} - 1\right] + \frac{1}{(1+x)^3 \ln 2 \log_2\left(1+\frac{1}{x}\right)}\left[\frac{1}{x \ln 2 \log_2\left(1+\frac{1}{x}\right)} - 1\right] \cdot \frac{1}{x^2 \log_2\left(1+\frac{1}{x}\right)} \quad (50)$$

Since $\left[\frac{1}{(1+x)\ln 2 \log_2\left(1+\frac{1}{x}\right)} - 2\right] < \left[\frac{1}{(1+x)\ln 2 \log_2\left(1+\frac{1}{x}\right)} - 1\right]$, we have:

$$f''(x) > \frac{1}{x^3 \log_2\left(1+\frac{1}{x}\right)}\left[\frac{1}{(1+x)\ln 2 \log_2\left(1+\frac{1}{x}\right)} - 2\right]^2 + \frac{1}{(1+x)^3 \ln 2 \log_2\left(1+\frac{1}{x}\right)}\left[\frac{1}{x \ln 2 \log_2\left(1+\frac{1}{x}\right)} - 1\right] \cdot \frac{1}{x^2 \log_2\left(1+\frac{1}{x}\right)} \quad (51)$$

Thus, if $g(x) = \frac{1}{x \ln 2 \log_2\left(1+\frac{1}{x}\right)} - 1 > 0$, the function $f(x)$ is convex. Moreover, $g(x) > 0$ equals to:

$$\frac{1}{x \ln 2 \log_2\left(1+\frac{1}{x}\right)} > 1 \quad (52)$$

$$\log_2\left(1+\frac{1}{x}\right) < \frac{1}{x \ln 2} \quad (53)$$

Based on the Bottoming formula, we have:
$$\log_2\left(1+\frac{1}{x}\right) < \log_2 e^{\frac{1}{x}} \quad (54)$$

which equals to:
$$\left(1+\frac{1}{x}\right) < e^{\frac{1}{x}} \quad (55)$$

Since the Taylor expansion of $e^{\frac{1}{x}}$ is:
$$e^{\frac{1}{x}} = \sum_{n=0}^{\infty} \frac{(1/x)^n}{n!} \quad (56)$$

which is larger than $\left(1+\frac{1}{x}\right)$. Thus, the $f(x)$ is convex. Similar to the L-Lipschitz smooth, the $y(x)$, $g(x)$, $h(x)$, and $z(x)$ are all convex.

Additionally, based on (22), the terms of $\Gamma_{s_i}$ relate to $r_i$ are:
$$\Gamma_{s_i}(r_i) = \sum_{i=1}^{U} \omega_T^i \left(\frac{f_e^i}{\lambda(r_i)c_{min}}\right) + \sum_{i=1}^{U} \omega_E^i \left(\xi_e(\lambda(r_i)c_{min})^2 \varphi_e f_e^i\right) \quad (57)$$

For (57), we assume that the $\lambda(r_i)$ is L-Lipschitz smooth and convex, thus, for $\Gamma_{s_i}(r_i)$, it is also L-Lipschitz smooth and convex. Therefore, the Corollary 2 holds. ∎

**Corollary 3.** The average complex of the Li-GD algorithm is $O(X\overline{K}\mathcal{F}Mx^3 \ln^2(x))$, where $\overline{K} = \frac{\sum_{i=1}^{\mathcal{F}} K^i}{\mathcal{F}}$

*Proof.* Based on Corollary 2, the convergence time of Li-GD approach is $K = \frac{\|x^0 - x^*\|_2^2}{2\eta\epsilon}$. In Li-GD algorithm, for each round of iteration, we need to calculate the gradient for $B_i$ and $r_i$, and the number of mobile users is $U$. Moreover, the gradient calculation of $B_i$, which is shown in (35), is much more complex than $r_i$. The (35) can be rewritten as:

$$f'(x) = \frac{1}{x^2 \log_2\left(1+\frac{1}{x}\right)}\left(\frac{1}{(1+x)\ln 2 \log_2\left(1+\frac{1}{x}\right)} - 1\right)$$
$$= \frac{1}{x^2 \log_2\left(1+\frac{1}{x}\right)} \cdot \frac{1}{(1+x)\ln 2 \log_2\left(1+\frac{1}{x}\right)} - \frac{1}{x^2 \log_2\left(1+\frac{1}{x}\right)} \quad (58)$$

Since the first term is more complex than the second term, we only consider the first term, let:
$$h'(x) = \frac{1}{x^2 \log_2\left(1+\frac{1}{x}\right)} \cdot \frac{1}{(1+x)\ln 2 \log_2\left(1+\frac{1}{x}\right)}$$
$$= \frac{1}{x^2(1+x)\left[\log_2\left(1+\frac{1}{x}\right)\right]^2 \ln 2} = O(x^3 \ln^2(x)) \quad (59)$$

Moreover, for each mobile user in each round of GD algorithm, the (59) should be calculated. Additionally, the number of layers in the Li-GD algorithm is $\mathcal{F}$. Thus, the worst case complexity of the Li-GD algorithm is $O(XK^*\mathcal{F}Mx^3 \ln^2(x))$, where $K^*$ is the maximum convergence time during $\mathcal{F}$ layers and $K^* = \max\{K^1, K^2, ..., K^{\mathcal{F}}\}$, $M$ is the number of subchannels. The average complexity of the Li-GD algorithm is $O(X\overline{K}\mathcal{F}Mx^3 \ln^2(x))$, where $\overline{K} = \frac{\sum_{i=1}^{\mathcal{F}} K^i}{\mathcal{F}}$. ∎

**Corollary 4.** The Li-GD algorithm can accelerate the convergence of GD algorithm while reducing complexity.

*Proof.* From Corollary 2 and Corollary 3, we can conclude that for the fixed precision $\varepsilon$ and step size $\eta$, the convergence time $K$ is corelated to the starting point $\boldsymbol{B}^{(0)}$, $\boldsymbol{P}^{(0)}$, and $\boldsymbol{r}^{(0)}$, and the complexity is associated with the convergence time $K$. Thus, for reducing convergence time and complexity, the starting point should be selected carefully.

In the traditional GD algorithm, for each round of GD, the starting values are $\boldsymbol{B}^{(0)}$, $\boldsymbol{P}^{(0)}$, and $\boldsymbol{r}^{(0)}$. The convergence time is $K_1 = \max\left\{\frac{\|\boldsymbol{B}^{(0)} - \boldsymbol{B}^*\|_2^2}{2\eta\epsilon}, \frac{\|\boldsymbol{r}^{(0)} - \boldsymbol{r}^*\|_2^2}{2\eta\epsilon}, \frac{\|\boldsymbol{P}^{(0)} - \boldsymbol{P}^*\|_2^2}{2\eta\epsilon}\right\}$. However, as introduced in Section 4.1, for the Li-GD algorithm, the starting values are the optimal solutions of last layer, which are $\boldsymbol{B}^{j+1(0)} = \boldsymbol{B}^{j*}$, $\boldsymbol{P}^{j+1(0)} = \boldsymbol{P}^{j*}$, and $\boldsymbol{r}^{j+1(0)} = \boldsymbol{r}^{j*}$. Therefore, the convergence time of Li-GD algorithm is $K_2 = \max\left\{\frac{\|\boldsymbol{B}^{j+1(0)} - \boldsymbol{B}^{j*}\|_2^2}{2\eta\epsilon}, \frac{\|\boldsymbol{r}^{j+1(0)} - \boldsymbol{r}^{j*}\|_2^2}{2\eta\epsilon}, \frac{\|\boldsymbol{P}^{j+1(0)} - \boldsymbol{P}^{j*}\|_2^2}{2\eta\epsilon}\right\}$. Since $|\boldsymbol{B}^{j+1(0)} - \boldsymbol{B}^{j*}|$, $|\boldsymbol{P}^{j+1(0)} - \boldsymbol{P}^{j*}|$, and $|\boldsymbol{r}^{j+1(0)} - \boldsymbol{r}^{j*}|$ are much smaller than $|\boldsymbol{B}^{(0)} - \boldsymbol{B}^*|$, $|\boldsymbol{P}^{(0)} - \boldsymbol{P}^*|$, and $|\boldsymbol{r}^{(0)} - \boldsymbol{r}^*|$, the convergence time is accelerated.

Additionally, there are $\mathcal{F}$ layers, for the traditional GD algorithm, the total convergence time is $\mathcal{F}K_1$. For Li-GD algorithm, the total convergence time is $K_1 + \sum_{j=2}^{\mathcal{F}} K_2^j$. Since $K_2^j$ is much smaller than $K_1$, the complexity is declined. ∎

**Corollary 5.** The approximate error $\varphi$ of Li-GD algorithm is less than $\frac{\varepsilon}{\rho_{min}(1-B_{max})\log_2\left(1+\frac{P_{min}}{\Delta^* + \frac{\alpha P_{max}}{2}}\right)}$.

*Proof.* As shown in Corollary 1, the approximation is caused by the approximate of $\beta_{n,i}^m \in \{0,1\}$ and $\beta_{j,i}^k \in \{0,1\}$ to $\beta_{n,i}^m \in [0\ 1]$ and $\beta_{j,i}^k \in [0\ 1]$. Therefore, according to (7), we have:

$$\partial = \frac{1}{\rho_2(1-B_i^2)\log_2\left(1+\frac{P_2}{\mathbb{R}^*+\mathbb{R}_{B_i^2}'}\right)} - \frac{1}{\rho_1 B_i^1 \log_2\left(1+\frac{P_1}{\mathbb{R}^*+\mathbb{R}_{B_i^1}'}\right)} \quad (60)$$

where $B_i^2$ means that the value of $B_i$ is larger than 0.5, $\rho_2$ is the probability that $B_i > 0.5$; $B_i^1$ means that the value of $B_i$ is smaller than 0.5; $\rho_1$ is the probability that $B_i^1 < 0.5$; $\mathbb{R}^*$ is the inter-cell interference and intra-cell interference under optimal circumstances; $\mathbb{R}_{B_i^2}'$ and $\mathbb{R}_{B_i^1}'$ are the increased intra-cell interference and inter-cell interference under these two circumstances. Additionally, $0 < \rho_2 < 1$, $0.5 < B_2 < 1$, $0 < \rho_1 < 1$, and $0 < B_1 < 0.5$. Let $P_2 = \alpha P_1$, therefore, the (60) equals to:

$$\partial < \frac{1}{\rho_2(1-B_i^2)\log_2\left(1+\frac{P_2}{\mathbb{R}^*+\mathbb{R}_{B_i^2}'}\right)} - \frac{2}{\log_2\left(1+\frac{P_1}{\mathbb{R}^*+\mathbb{R}_{B_i^1}'}\right)} \quad (61)$$

Moreover, for the $\mathbb{R}_{B_i^2}'$ and $\mathbb{R}_{B_i^1}'$, we have:

$$\mathbb{R}_{B_i}' = \rho_2(1-B_i^2)P_2 - \rho_1 B_i^1 P_1 < \frac{\rho_2 P_2 - \rho_1 P_1}{2}$$
$$= \frac{P_1(\alpha\rho_2-\rho_1)}{2} < \frac{P_{max}(\alpha\rho_2-\rho_1)}{2} < \frac{\alpha P_{max}}{2} \quad (62)$$

Thus, (62) equals to:

$$\partial < \frac{1}{\rho_2(1-B_i^2)\log_2\left(1+\frac{P_2}{\Delta^*+\frac{\alpha P_{max}}{2}}\right)} - \frac{2}{\log_2\left(1+\frac{P_1}{\Delta^*+\frac{\alpha P_{max}}{2}}\right)}$$
$$< \frac{1}{\rho_2(1-B_i^2)\log_2\left(1+\frac{P_{min}}{\Delta^*+\frac{\alpha P_{max}}{2}}\right)} < \frac{1}{\rho_{min}(1-B_i^{max})\log_2\left(1+\frac{P_{min}}{\Delta^*+\frac{\alpha P_{max}}{2}}\right)} \quad (63)$$

Moreover, the accuracy of GD algorithm is $\varepsilon$, therefore, the approximate error of Li-GD algorithm is $\frac{\varepsilon}{\rho_{min}(1-B_i^{max})\log_2\left(1+\frac{P_{min}}{\Delta^*+\frac{\alpha P_{max}}{2}}\right)}$. ∎

Note that the convergence time can be reduced further by optimizing the step size or using self-adaptive step size. Moreover, lowering the accuracy can also accelerate the convergence. Therefore, by carefully achieving tradeoff between accuracy and convergence time also can improve the performance of Li-GD algorithm. However, these are not investigated in this paper.

## VI. PERFORMANCE EVALUATION

### A. Experimental Setup

*Network and Communication set.* We deploy 5 APs and 1250 users in the network. The system bandwidth is 10 MHz, which is available to all the APs. In addition, we assume that each subchannel can be accessed by at most 3 devices. The uplink channels are all independent and identically distributed Rayleigh fading channels. Refer to [32], the number of subchannels is 250; the maximum transmission power of device is 25dBm; the circuit power consumption of each edge server is 50dBm; the path loss exponent is 5; the noise power spectral density is -174dbm/Hz; the CPU cycles for 1bit task is $10^4$ cycles/bit.

*Dataset.* We use CIFAR-10 dataset in this paper. The CIFAR-10 dataset consists of 60000 32 × 32 RGB images in 10 classes (from 0 to 9), with 50000 training images and 1000 test images per class.

*DNN benchmarks.* There are many DNN models with different topologies have been proposed recently. For instance, NiN, tiny YOLOv2, VGG16, etc., are the well-known chain topology models; AlexNet, ResNet-18, etc., are the well-known DAG topology models. However, in this paper, we mainly evaluate the performance of the proposed algorithms on chain topology models, i.e., NiN (9 layers), YOLOv2 (17 layers) and VGG16 (24 layers).

*Evaluation benchmarks.* We compare the proposed algorithms against Device-Only (i.e., executing the entire DNN on the device), Edge-Only (i.e., executing the entire DNN on the edge), Neurosurgeon, and DNN surgeon. However, the DNN surgeon can operate on both chain topology models and DAG topology models. In this paper, since we mainly focus on chain topology, we only implement DNN surgeon on chain topology models (i.e., NiN, YOLOv2, and VGG16).

### B. Performance evaluation

In this section, we compare the performance of ECC-NOMA (ECC approach with NOMA channel) and ECC-OMA (ECC approach with OMA channel, shorted as ECC in the following sections) with Device-Only, Edge-Only, Neurosurgeon, and DNN surgeon. In this section, we use the Device-Only method as the baseline, i.e., the performance is normalized to the Device-Only method.

*Compare the ECC-NOMA and ECC with Device-Only and Edge-Only approach.* The performance of ECC-NOMA and ECC has been compared with Device-Only and Edge-Only strategies. The performance of latency speedup and energy consumption reduction is presented in Fig.2 and Fig.3, respectively.

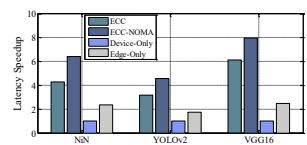 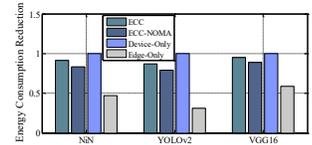

Fig.2. Latency speedup for different DNN models

Fig.3. Energy consumption reduction for different DNN models

From Fig.2, firstly, we can find that comparing with Device-Only approach, both the ECC and Edge-Only approach can reduce the inference latency. For instance, without NOMA, the latency speedup of ECC is 3.1 to 8 times higher than the Device-Only approach. This is because the whole inference task is executed on device in the Device-Only approach, in which the computing capability is lower than the edge server. Moreover, the performance of ECC approach is also better than that of the Edge-Only approach. The reason is that even the computing capability of edge server is better than the device, the large size of raw data makes the data transmission delay is high. Moreover, the performance of ECC with NOMA communication channel is better than that without NOMA channel. For instance, in NiN, the latency speedup of ECC without NOMA channel is 4.1 times while it is 6.1 times with NOMA channel. The performance of Device-Only approach is not affected by communication channel since it does not need to transmit data to edge server.

The performance of energy consumption reduction is presented in Fig.3. We can find that the energy consumption in both ECC and Edge-Only approach is larger than that in Device-Only approach. For instance, the energy consumption reduction of ECC is 0.85 to 0.97 times lower than the Device-Only approach. The reason is that in ECC and Edge-Only approach, on one hand, there is large amount data transmission between device and edge server, on the other hand, the power for data processing in edge server is higher than that in device. Moreover, considering that the transmission power in NOMA based channel is higher than that in OMA channel to avoid intra-cell interference, the energy consumption in ECC-NOMA is higher than that in ECC.

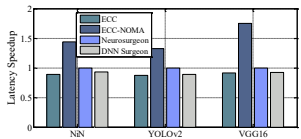

Fig.4. Latency speedup for different DNN models

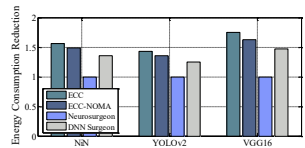

Fig.5. Energy consumption reduction for different DNN models

*Comparing ECC-NOMA and ECC with Neurosurgeon and DNN surgeon.* The performance of ECC-NOMA and ECC is compared with Neurosurgeon and DNN surgeon. The performance of latency speedup and energy consumption reduction is presented in Fig.4 and Fig.5, respectively. In this section, we use the Neurosurgeon as the baseline, i.e., the performance is normalized to the Neurosurgeon algorithm.

From Fig.4, we can find that the latency speedup in ECC and DNN surgeon is similar but a little worse than that in Neurosurgeon; however, the latency speedup in ECC-NOMA is much better than that in Neurosurgeon. This is because in ECC and DNN surgeon, the energy consumption and the limitation of computing resource in edge server is considered to find the optimal tradeoff between latency and energy consumption. Therefore, the computing resource that allocated in ECC and DNN surgeon is a little less than that in Neurosurgeon, which causes the reduction of latency speedup. When considering the NOMA channel in ECC, since the spectrum efficiency is improved, the data transmission rate is speedup. Therefore, the performance of ECC-NOMA is better than Neurosurgeon.

The performance of energy consumption reduction is presented in Fig.5. We can find that the energy consumption reduction in ECC is the best. The energy consumption reduction in DNN surgeon is better than that in Neurosurgeon. For instance, the energy consumption reduction in ECC is 1.5 to 1.7 times larger than Neurosurgeon, which is 1.3 to 1.49 times in DNN surgeon. This is because that the Neurosurgeon and DNN surgeon do not consider the energy consumption of system, but DNN surgeon considers the resource limitation of edge server; thus, the energy consumption in DNN surgeon is less than that in Neurosurgeon. Moreover, considering that the transmission power in NOMA channel is higher than that in OMA channel to avoid intra-cell interference, the energy consumption in ECC-NOMA is higher than that in ECC, but better than the other algorithms.

### C. Performance under different network conditions

In this section, we compare the performance of ECC-NOMA with Device-Only, Edge-Only, Neurosurgeon, and DNN surgeon under different network conditions, including different densities of mobile users (average number of users in each edge server), different number of subchannels, and different workloads. The results are presented in Fig.6 to Fig.11, respectively. In this section, we use the Device-Only method as the baseline, i.e., the performance is normalized to the Device-Only method.

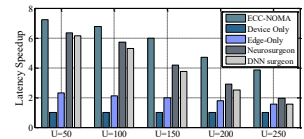

Fig.6. Latency speedup under different user densities

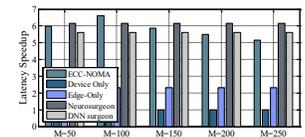

Fig.7. Latency speedup under different number of subchannels

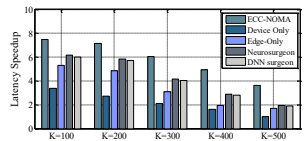

Fig.8. Latency speedup under different workload

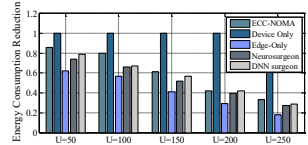

Fig.9. Energy consumption reduction under different user densities

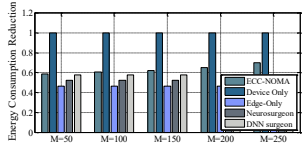

Fig.10. Energy consumption reduction under different number of subchannels

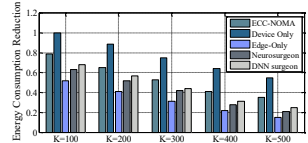

Fig.11. Energy consumption reduction under different workload

The latency of ECC-NOMA, Device-Only, Edge-Only, Neurosurgeon, and DNN surgeon under different user densities is presented in Fig.6. We can find that with the increasing of the user density, the latency increases in these algorithms except for Device-Only approach. Since in Device-Only approach, the whole inference model is in the mobile device, it is not affected by the variation of user density. The latency performance of ECC-NOMA is the best. The advantages of ECC becomes obviously with the increasing of the user density. For instance, when $U = 10$, the latency speedup of ECC-NOAM and Neurosurgeon is 7.1 and 6.3, respectively; when $U = 90$, this becomes 3.85 and 1.9, respectively. The reason is that the NOMA improve the spectrum efficiency, which can guarantee the data transmission rate under crowed network.

The performance of latency speedup under different number of subchannels is presented in Fig.7. We can find that with the increasing of the number of subchannels, the latency speedup of ECC-NOMA increases first then reducing after $M = 6$. This is because when the number of subchannels increases, the bandwidth of each subchannel reduces. Even the number of users in each subchannel reduces, if the bandwidth of each subchannel is small enough, the data transmission rate will be reduced seriously. The performance of Edge-Only, Neurosurgeon, and DNN surgeon are not affected because they do not use the NOMA channel. Moreover, when $M = 4$, the performance of ECC-NOMA is the best. For the other values of $M$, the performance of Neurosurgeon is the best. The reason is the same as that in Fig.6.

The performance of latency speedup under different workloads is presented in Fig.8, where $k$ means the average number of works in each mobile user. In Fig.8, with the increasing of workloads, the latency in mobile devices increases, too. Therefore, in this Section, we use the latency of mobile devices when $K = 500$ as the baseline. Therefore, with the increasing of workloads, the latency increases in both these five algorithms. The latency speedup of ECC-NOMA is better than the other algorithms. Moreover, the advantage of ECC-NOMA becomes obviously with the increasing of workload. The reasons are similar with that in Fig.6 and Fig.7.

The energy consumption of ECC-NOMA, Device-Only, Edge-Only, Neurosurgeon, and DNN surgeon under different user densities is presented in Fig.9. We can find that with the increasing of the user density, the energy consumption increases in these algorithms except for Device-Only approach. Since in Device-Only approach, the whole inference model is in the mobile device, its energy consumption is not affected by the variation of user density. The energy consumption reduction of ECC-NOMA is the best. The advantage of ECC-NOMA becomes smaller with the increasing of the user density. This is because with the increasing of the user density, more users will share the same subchannel, for maintain high data transmission rate, the transmission power should be improved in NOMA.

The energy consumption reduction under different number of subchannels is presented in Fig.10. We can find that with the increasing of the number of subchannels, the energy consumption of ECC-NOMA reduces slightly. This is because when the number of subchannels increases, the number of mobile users in each subchannel reduces. Then the intra-cell interference reduces. Therefore, the user and edge server can lower their transmission power. However, the increasing of the number of subchannels will cause the increasing of latency, which will contribute to the increasing of energy consumption. Thus, the energy consumption reduction in ECC-NOMA is slight. The performance of Edge-Only, Neurosurgeon, and DNN surgeon are not affected because they do not use the NOMA channel. Moreover, the performance of Device-Only is the best, the reasons are the same with that in Fig.3.

The performance of energy consumption under different workloads is presented in Fig.11. In Fig.11, with the increasing of workloads, the energy consumption in mobile devices increases, too. Therefore, in this figure, we use the latency of mobile devices when $K = 500$ as the baseline. Therefore, with the increasing of workloads, the energy consumption increases in both these five algorithms. The energy consumption reduction of ECC-NOMA is better than the other algorithms. Moreover, the advantage of ECC-NOMA becomes obviously with the increasing of workload. The reasons are similar with that in Fig.8.

## VII. CONCLUSION

In this paper, for accelerating the model inference at edge, we integrate the NOMA into split inference in MEC, and propose the effective communication and computing resource allocation algorithm, shorted as ECC. Specifically, when the mobile user has a large model inference task needed to be calculated in the NOMA-based MEC, it will take the energy consumption of both device and edge server and the inference latency into account to find the optimal model split strategy, subchannel allocation strategy, and transmission power allocation strategy. Since the minimum inference delay and energy consumption cannot be satisfied simultaneously, and the variables of sub-channel allocation and model split are discrete, the gradient descent (GD) algorithm is adopted to find the optimal tradeoff between them. Moreover, the loop iteration GD approach (Li-GD) is proposed to reduce the complexity of GD algorithm that caused by the discrete of model segmentation. Additionally, the properties of the proposed algorithms are also investigated, which demonstrate the effectiveness of the proposed algorithms.

### ACKNOWLEDGMENT

This work was supported in part by a grant from NSFC Grant no. 62101159, NSF of Shandong Grant no. ZR2021MF055, and also the Research Grants Council of Hong Kong under the Areas of Excellence scheme grant AoE/E-601/22-R.

### REFERENCES

[1] S. Zeng, Z. Li, H. Yu, Zh. Zhang, L. Luo, B. Li, D. Niyato, "HFedMS: Heterogeneous federated learning with memorable data semantics in industrial metaverse," IEEE Transactions on Cloud Computing, no.11, vol.3, 2023, pp:

[2] M. Xu, D. Niyato, J. Chen, H. Zhang, J. Kang, Z. Xiong, Sh. Mao, Zh. Han, "Generative AI-empowered simulation for autonomous driving in vehicular mixed reality metaverses,", IEEE Journal of Selected Topics in Signal Processing, Early Access, 2023, pp:

[3] A. Ndikumana, H.T. Nguyen, D.H. Kim, K.T. Kim, C.S. Hong, "Deep learning-based caching for self-driving cars in multi-access edge computing," IEEE Transactions on Intelligent Transportation Systems, vol.22, no.5, 2021, pp:

[4] H. Tian, X. Xu, L. Qi, X. Zhang, W. Dou, Sh. Yu, Q. Ni, "CoPace: Edge computation offloading and caching for self-driving with deep reinforcement learning," IEEE Transactions on Vehicular Technology, no.70, no1.12, 2021, pp:

[5] S. Yin, Ch. Fu, S. Zhao, K. Li, X. Sun, T. Xu, E. Chen, "A survey on multimodal large language models," [online], 2023, available: https://arxiv.org/pdf/2306.13549.pdf.

[6] Y. Huang, X. Qiao, W. Lai, S. Dustdar, J. Zhang, and J. Li, "Enabling DNN acceleration with data and model parallelization over ubiquitous end devices," IEEE Internet Things J., vol. 9, no. 16, pp. 15053–15065, Aug. 2022.

[7] L. Zeng, X. Chen, Z. Zhou, L. Yang, and J. Zhang, "CoEdge: Cooperative DNN inference with adaptive workload partitioning over heterogeneous edge devices," IEEE/ACM Trans. Netw., vol. 29, no. 2, pp. 595–608, Apr. 2021.

[8] S. Zhang, S. Zhang, Z. Qian, J. Wu, Y. Jin, and S. Lu, "DeepSlicing: Collaborative and adaptive CNN inference with low latency," IEEE Trans. Parallel Distrib. Syst., vol. 32, no. 9, pp. 2175–2187, Sep. 2021.

[9] F. Dong, H. Wang, D. Shen, Z. Huang, Q. He, J. Zhang, L. Wen, T. Zhang, "Multi-exit DNN inference acceleration based on multi-dimensional optimization for edge intelligence," IEEE Transactions on Mobile Computing, no.22, vol.9, 2023, pp: 5389-5405.

[10] C. Yang, J.J. Kuo, J.P. Sheu, K.J. Zheng, "Cooperative distributed deep neural network deployment with edge computing," IEEE ICC, 2021, Montreal, Canada, pp: 1-6.

[11] T. Mohammed, C.J. Wong, R. Babbar, M.D. Francesco, "Distributed inference acceleration with adaptive DNN partitioning and offloading," IEEE INFOCOM, 2020, Toronto, Canada, pp: 1-10.

[12] A.E. Eshratifar, M.S. Abrishami, M. Pedram, "JointDNN: An efficient training and inference engine for intelligent mobile cloud computing services," IEEE Transactions on Mobile Computing, vol.20, no.2, 2021, pp: 565-576.

[13] M. Zhou, B. Zhou, H. Wang, F. Dong, W. Zhao, "Dynamic path based DNN synergistic inference acceleration in edge computing environment," IEEE ICPADS, 2021, Beijing, China, pp: 1-8.

[14] H. Liang, Q. Sang, Ch. Hu, D. Cheng, X. Zhou, D. Wang, W. Bao, Y. Wang, "DNN surgery: Accelerating DNN inference on the edge through layer partitioning," IEEE transactions on Cloud Computing, vol.11, no.3, 2023, pp: 3111-3125.

[15] X. Tang, X. Chen, L. Zeng, Sh. Yu, L. Chen, "Joint multiuser DNN partitioning and computational resource allocation for collaborative edge intelligence," IEEE Internet of Things Journal, vol.8, no.12, 2021, pp: 9511-9522.

[16] E. Li, L. Zeng, Zh. Zhou, X. Chen, "Edge AI: On-demand accelerating deep neural network inference via edge computing," IEEE Transactions on Wireless Communication, vol.19, no.1, 2020, pp: 447-457.

[17] Z. Ding, P. Fan, and H. V. Poor, "Impact of user pairing on 5G nonorthogonal multiple-access downlink transmissions," IEEE Trans.

Veh. Technol., vol. 65, no. 8, pp. 6010–6023, Aug. 2016.
[18] M. Vaezi, R. Schober, Z. Ding, and H. V. Poor, "Non-orthogonal multiple access: Common myths and critical questions," IEEE Wireless Commun., vol. 26, no. 5, pp. 174–180, Oct. 2019.
[19] Z. Ding, R. Schober, P. Fan, and H. V. Poor, "Simple semi-grant free transmission strategies assisted by non-orthogonal multiple access," IEEE Trans. Commun., vol. 67, no. 6, pp. 4464–4478, Jun. 2019.
[20] Z. Ding, P. Fan, G. K. Karagiannidis, R. Schober, and H. V. Poor, "NOMA assisted wireless caching: Strategies and performance analysis," IEEE Trans. Commun., vol. 66, no. 10, pp. 4854–4876, Oct. 2018.
[21] A. E. Mostafa, Y. Zhou, and V. W. S. Wong, "Connection density maximization of narrowband IoT systems with NOMA," IEEE Trans. Wireless Commun., vol. 18, no. 10, pp. 4708–4722, Oct. 2019.
[22] D. Zhai, R. Zhang, L. Cai, and F. R. Yu, "Delay minimization for massive Internet of Things with non-orthogonal multiple access," IEEE J. Sel. Topics Signal Process., vol. 13, no. 3, pp. 553–566, Jun. 2019.
[23] A. Kiani and N. Ansari, "Edge computing aware NOMA for 5G networks," IEEE Internet Things J., vol. 5, no. 2, pp. 1299–1306, Apr. 2018.
[24] L. P. Qian, A. Feng, Y. Huang, Y. Wu, B. Ji, and Z. Shi, "Optimal SIC ordering and computation resource allocation in MEC-aware NOMA NB-IoT networks," IEEE Internet Things J., vol. 6, no. 2, pp. 2806–2816, Apr. 2019.
[25] Y. Wu, K. Ni, C. Zhang, L. P. Qian, and D. H. K. Tsang, "NOMA assisted multi-access mobile edge computing: A joint optimization of computation offloading and time allocation," IEEE Trans. Veh. Technol., vol. 67, no. 12, pp. 12244–12258, Dec. 2018.
[26] X. Diao, J. Zheng, Y. Wu, and Y. Cai, "Joint computing resource, power, and channel allocations for D2D-assisted and NOMA-based mobile edge computing," IEEE Access, vol. 7, pp. 9243–9257, 2019.
[27] Z. Song, Y. Liu, and X. Sun, "Joint radio and computational resource allocation for NOMA-based mobile edge computing in heterogeneous networks," IEEE Commun. Lett., vol. 22, no. 12, pp. 2559–2562, Dec. 2018.
[28] M. Peng, K. Zhang, J. Jiang, J. Wang, and W. Wang, "Energy efficient resource assignment and power allocation in heterogeneous cloud radio access networks," IEEE Trans. Veh. Technol., vol. 64, no. 11, pp. 5275–5287, Nov. 2015.
[29] G. Cui, Q. He, X. Xia, F. Chen, F. Dong, H. Jin, Y. Yang, "OL-EUA: Online user allocation for NOMA-based mobile edge computing," IEEE Transactions on Mobile Computing, vol.22, no.4, 2023, pp. 2295-2306.
[30] P. Lai, Q. He, F. Chen, M. Abdelrazek, J. Hosking, J. Grundy, Y. Yang, "Online user and power allocation in dynamic NOMA-based mobile edge computing," IEEE Transactions on Mobile Computing, vol.22, no.11, 2023, pp. 6676-6689.
[31] G. Cui, Q. He, X. Xia, F. Chen, T. Gu, H. Jin, Y. Yang, "Demand response in NOMA-based mobile edge computing: A two-phase game-theoretical approach," IEEE Transactions on Mobile Computing, vol.22, no.3, 2023, pp. 1449-1463.
[32] B. Liu, Ch. Liu, M. Peng, "Resource allocation for energy-efficient MEC in NOMA-enabled massive IoT Networks," IEEE Journal on Selected Aeras in Communications, vol.39, no.4, 2021, pp: 1015-1027.
[33] S. Rostami, K. Heiska, O. Puchko, K. Leppanen, M. Valkama, "Novel wake-up scheme for energy-efficient low-latency mobile devices in 5G networks," IEEE Transactions on Mobile Computing, vol.21, no.4, 2021, pp:
[34] D. Huang, P. Wang, D. Niyato, "A dynamic offloading algorithm for mobile computing," IEEE Transactions on Wireless Communication, vol.11, no.6, 2012, pp:
[35] A.M. Groba, P.J. Lobo, M. Chavarrias, "QoE-Aware dual control system to guarantee battery lifetime for mobile video applications," IEEE Transactions on Consumer Electronics, vol.65, no.4, 2019, pp:
[36] J. Lu, Q. Li, B. Guo, J. Li, Y. Shen, G. Li, and H. Su, "A multi-task-oriented framework for mobile computation offloading," IEEE Transactions on Cloud Computing, 2019.
[37] L. Lin, X. Liao, H. Jin, and P. Li, "Computation offloading toward edge computing," Proceedings of the IEEE, vol. 107, no. 8, pp. 1584–1607, 2019.
[38] Y. Kang, J. Hauswald, C. Gao, A. Rovinski, T. Mudge, J. Mars, and L. Tang, "Neurosurgeon: Collaborative intelligence between the cloud and mobile edge," in Proc. ACM ASPLOS'17, Xi'an, China, Apr. 2017.
[39] E. Cuervo, A. Balasubramanian, D.-k. Cho, A. Wolman, S. Saroiu, R. Chandra, and P. Bahl, "Maui: making smartphones last longer with code offload," in Proc. ACM MobiSys'10, San Francisco, CA, Jun. 2010.
[40] S. Teerapittayanon, B. McDanel, and H. Kung, "Distributed deep neural networks over the cloud, the edge and end devices," in Proc. IEEE ICDCS'17, Atlanta, GA, Jun. 2017.
[41] J. Mao, Z. Yang, W. Wen, C. Wu, L. Song, K. W. Nixon, X. Chen, H. Li, and Y. Chen, "MeDNN: A distributed mobile system with enhanced partition and deployment for large-scale DNNs," in 2017 IEEE/ACM International Conference on Computer-Aided Design (ICCAD). IEEE, 2017, pp. 751–756.
[42] T. Mohammed, C. Joe-Wong, R. Babbar, and M. D. Francesco, "Distributed inference acceleration with adaptive DNN partitioning and offloading," in IEEE INFOCOM 2020 - IEEE Conference on Computer Communications, 2020, pp. 854–863.
[43] Z. Zhao, K. M. Barijough, and A. Gerstlauer, "Deepthings: Distributed adaptive deep learning inference on resource-constrained IoT edge clusters," IEEE Transactions on Computer-Aided Design of Integrated Circuits and Systems, vol. 37, no. 11, pp. 2348–2359, 2018.
[44] X. Tang, X. Chen, L. Zeng, S. Yu, and L. Chen, "Joint multiuser DNN partitioning and computational resource allocation for collaborative edge intelligence," IEEE Internet of Things Journal, 2020.
[45] H. Li, C. Hu, J. Jiang, Z. Wang, Y. Wen, and W. Zhu, "Jalad: Joint accuracy-and latency-aware deep structure decoupling for edge cloud execution," in 2018 IEEE 24th International Conference on Parallel and Distributed Systems (ICPADS). IEEE, 2018, pp. 671–678.
[46] J. H. Ko, T. Na, M. F. Amir, and S. Mukhopadhyay, "Edge-host partitioning of deep neural networks with feature space encoding for resource-constrained internet-of-things platforms," in 2018 15th IEEE International Conference on Advanced Video and Signal Based Surveillance (AVSS). IEEE, 2018, pp. 1–6.
[47] Z. Huang, F. Dong, D. Shen, J. Zhang, H. Wang, G. Cai, and Q. He, "Enabling low latency edge intelligence based on multi-exit DNNs in the wild," in 2021 IEEE 41st International Conference on Distributed Computing Systems (ICDCS), 2021, pp. 729–739.
[48] Q. Sun, Y. Zhang, S. Jin, J. Wang, X. Gao, and K.-K. Wong, "Downlink massive distributed antenna systems scheduling," IET Commun., vol. 9, no. 7, pp. 1006–1016, May 2015.
[49] Y. Saito, Y. Kishiyama, A. Benjebbour, T. Nakamura, A. Li, and K. Higuchi, "Non-orthogonal multiple access (NOMA) for cellular future radio access," in Proc. IEEE Veh. Technol. Conf., 2013, pp. 1–5.
[50] K. Wang, Y. Liu, Z. Ding, A. Nallanathan, and M. Peng, "User association and power allocation for multi-cell non-orthogonal multiple access networks," IEEE Trans. Wireless Commun., vol. 18, no. 11, pp. 5284–5298, Nov. 2019.
[51] S. Bubeck, Convex Optimization: Algorithms and Complexity. Now Foundations and Trends, 2015, USA.
[52] W.Q. Ren, Y.B. Qu, Y.Q. Jing, C. Dong, H. Sun, Q.H. Wu, S. Guo, "A survey on collaborative DNN inference for edge intelligence," Machine Intelligence Research, vol.2023, no.20, 2023, pp: 370-395.
[53] N. Li, X. Yuan, Zh. Zhang, "CoTree: Region-free and decentralized edge server cooperation," ACM HPDC workshop on FIRME, Minneapolis, USA, 2022.
[54] Z. Ning, H. Hu, X. Wang, L. Guo, S. Guo, G. Wang, X. Gao, "Mobile edge computing and machine learning in the internet of Unmanned Aerial Vehicles: A survey," ACM Computing Surveys, vol.56, no.1, 2023, pp: 1-31.